\documentclass[12pt,twoside,float,german]{article}
\usepackage{amsmath}
\usepackage{amssymb}
\usepackage{epsfig}
\newcommand{\be}{\begin{equation}}
\newcommand{\ee}{\end{equation}}
\newcommand{\bea}{\begin{eqnarray}}
\newcommand{\eea}{\end{eqnarray}}
\newcommand{\nn}{\nonumber}
\newcommand{\nnl}{\nonumber\\}

\begin{document}
%--------------------------------------------------------------------

\title{Nonequilibrium Quantum-Field Dynamics and Off-Shell Transport
for $\phi^4$-theory in 2+1 dimensions\footnote{Part of the PhD
thesis of S. Juchem}}

\author{S. Juchem$^1$,
W. Cassing$^1$ and C. Greiner$^2$\\[5mm] {\normalsize $^1$
Institut f\"{u}r Theoretische Physik,
 Universit\"{a}t Giessen}\\
    {\normalsize D-35392 Giessen, Germany} \\
   {\normalsize  Institut f\"{u}r Theoretische Physik,
 Universit\"{a}t Frankfurt}\\
    {\normalsize D-60054 Frankfurt, Germany}}
%\date{ }
\maketitle

\begin{abstract}
We solve the Kadanoff-Baym equations for nonequilibrium initial
configurations of the $\phi^4$-theory in 2+1 dimensions and
compare to explicit solutions of generalized transport equations
for the same theory. The latter transport equations are derived
from the Kadanoff-Baym equation in a first order gradient
expansion in phase space and explicitly retain the off-shell
dynamics as inherent in the time-dependent spectral functions. The
solutions of these equations compare very well with the exact
solutions of the full Kadanoff-Baym equations with respect to the
occupation numbers of the individual modes, the spectral evolution
as well as the chemical equilibration process. Furthermore, the
proper equilibrium off-shell distribution is reached for large
times contrary to the quasiparticle Boltzmann limit. We
additionally present a direct comparison of the solution of the
generalized transport equations in the Kadanoff-Baym and
Botermans-Malfliet form; both solutions are found to agree very
well with each other.
\end{abstract}

\noindent \vspace{10mm} \noindent PACS:  06.60.+w; 05.70.Ln;
11.10.Wx; 24.10.-i; 25.75.-q


\newpage
%\narrowtext

\section{\label{sec:intro} Introduction}

Nonequilibrium many-body theory or quantum-field theory has become
a major topic of research for transport processes in nuclear
physics, in cosmological particle physics as well as condensed
matter physics. In particular the dynamics of heavy-ion collisions
at various bombarding energies -- as representative of complex
systems with strongly interacting particles or fields -- has
always been a major motivation for research on nonequilibrium
quantum many-body physics. The initial state of such collisions
represents an extreme nonequilibrium situation, whereas the final
state might exhibit a certain degree of thermalization in
central reactions. Indeed, at the presently highest energy
heavy-ion collider experiments at the Relativistic Heavy-Ion
Collider (RHIC), where one expects to create experimentally an
intermediate state of deconfined  matter (denoted as quark-gluon
plasma (QGP)), there are experimental signatures for an early
thermalization accompanied with the generation of a very large
pressure. Such a rapid thermalization cannot be explained by
binary on-shell parton scattering using cross sections from
perturbative QCD (pQCD) \cite{BerndM}. Presently open
questions in high-energy nuclear physics are, if i) the quantum
off-shell equilibration proceeds much faster than ordinary
two-body on-shell Boltzmann kinetics and if ii) multi-parton
interactions play a decisive role.

We here primarily focus on the
first issue and present solutions for the full quantum time
evolution of nonequilibrium $\phi^4$-theory in 2+1 dimensions in
comparison to generalized off-shell transport equations -- derived
in the last years \cite{caju1,caju2} -- as well as the on-shell
quasiparticle Boltzmann limit. The scalar $\phi^4$-theory, of
cause, is much simpler than a non-abelian gauge-field theory like
QCD, however, provides a suitable testing ground for approximation
schemes to the field theoretical many-body problem.

We recall that a proper basis for  nonequilibrium quantum
many-body physics is provided by the closed-time-path (CTP)
functional integral (`Schwinger-Keldysh'
\cite{Sc61,BM63,Ke64,Cr68,Ch85}) formalism. The essential merit of
this real-time Green function technique is its applicability to
all kind of quantum nonequilibrium phenomena.  In this context the
most prominent role is played by the  causal Dyson-Schwinger
equations of motion for the one-particle Green functions, i.e. the
Kadanoff-Baym (KB) equations \cite{KB,DuBois,dan84a,calhu,Haug,GL98a}.
The latter, furthermore, have served as a basis for the derivation
of generalized transport equations, which  play the central role
in almost all practical simulations of nonrelativistic and
relativistic  nonequilibrium many-body systems. In the past, a
major impetus has been to derive semiclassical Boltzmann-like
transport equations within the standard quasiparticle
approximation \cite{Ca77,Ca78,Ca90,CMO}. More recently, off-shell
extensions by means of a gradient expansion in the space-time
inhomogeneities have been formulated for several systems: for a
relativistic electron-photon plasma \cite{BB72}, for transport of
electrons in a metal with external electrical field \cite{LSV86}
and in semiconductors \cite{Haug,SL94}, for transport of nucleons
in heavy-ion reactions at intermediate energies  \cite{botmal},
for transport of particles in $\phi^4$-theory \cite{calhu,danmrow}
as well as for transport of partons in high-energy heavy-ion
reactions \cite{Ma95,Ge96,BD98,BI99}.

Within all these derivations the  particle self-energies --
entering the KB equations -- have to fulfill consistency relations
in order to preserve symmetries of the fundamental Lagrangian
\cite{KB,knoll1,knoll5}. In this respect an elegant way for
achieving selfconsistency is provided by the
two-particle-irreducible (2PI) effective action
\cite{calhu,2PIEA,CJT}. Including the background field and the
correlation function on the same footing, collisions can be
naturally included by taking into account contributions beyond the
leading mean-field order of the generating functional. Due to the
construction in terms of full Green functions this method ensures
the validity of all essential conservation laws.

Moreover, in the nuclear physics context the quantum dynamics of
the spectral function is a lively discussed issue in the
microscopic modeling of hadronic resonances with a broad mass
distribution. This is of particular relevance for simulations of
heavy-ion reactions where e.g. the $\Delta $-resonance or the
$\rho $-meson already show a large decay width in vacuum.
Especially the $\rho $ vector meson is a promising hadronic
particle for showing possible in-medium modifications in hot and
compressed nuclear matter (see e.g. \cite{Wolf93,RW00,CB99}),
since the leptonic decay products   only  interact
electromagnetically.  Hence, a consistent formulation for the
transport of extremely short-lived particles beyond the standard
quasiparticle approximation is needed. On the one side, there
exist purely formal developments starting from a first order
gradient expansion of the underlying KB equations
\cite{KB,HHab,knoll3,knoll6,Leupold}, while on the other side
already first practical realizations for various questions have
emerged \cite{caju1,caju2,EBM99,caju3,cas03}.  However, a fully
{\it ab initio} investigation for a strongly interacting field
theoretical system in 3+1 dimensions --without any further
approximations -- does not exist so far as well as a rigorous
demonstration of the reliability of the different schemes for off-shell
transport dynamics.

Although the Kadanoff-Baym equation has been of wide interest,
(numerical) investigations of its full solution are rather scarce
for strongly interacting systems. First studies in the nuclear
physics context have been performed by Danielewicz \cite{dan84b},
who investigated the initial distribution of two occupied
non-relativistic Fermi spheres in momentum space for spatially
homogeneous systems. In comparison to a standard on-shell
semiclassical Boltzmann equation the full quantum Kadanoff-Baym
solution showed
 a larger collective relaxation time which
 was attributed to quantum
interference and off-shell effects. Similar quantum modifications
in the equilibration and momentum relaxation have been found in
\cite{Ca77} and for a relativistic situation in Ref.
\cite{CGreiner}. Furthermore, the spectral properties of
nuclear systems in the nonrelativistic regime have been first
examined on the basis of the KB equations in Refs.
\cite{koe1,koe2}. We mention that solutions of quantum transport
equations for semiconductors \cite{Haug,WJ99} -- to explore
relaxation phenomena on short time and distance scales -- have
become also a very active field of research (cf. Ref.
\cite{morabuch} for a recent overview).

The basic aim of our present study is to examine exact solutions of
the KB equations for strongly interacting (scalar) fields in
comparison to the solution of generalized kinetic equations as
obtained from a gradient expansion of the full Kadanoff-Baym
equation. Although such generalized transport equations have
already been used in the past, a check against the full solution
is still lacking. To this aim we will investigate the scalar $\phi^4$-theory
for the case of 2+1 space-time dimensions since -- first of all -- it
superceeds the rather artificial structure of the $\phi^4$-theory
in 1+1 dimensions, that has been investigated before. In the
latter case on-shell scattering processes are (col-)linear
and thus do not lead to an equilibration in accordance with an
appropriate change of the momentum distribution. Thermalization in
1+1 dimensions is only mediated by off-shell scattering
processes. This is a rather unusual situation, since it is known
from the investigation of heavy-ion collisions by means of quantum
kinetic theories, that binary on-shell scattering plays an
important role. In the 2+1 dimensional case both, on-shell and
off-shell collisions influence the equilibration behaviour. Thus
the theory in 2+1 dimensions is a well-suited laboratory in order
to investigate the differences between the full Kadanoff-Baym
theory and approximate schemes as given by the on-shell Boltzmann
limit or by generalized off-shell transport equations.
Furthermore, the
divergence structure of the $\phi^4$-theory in 2+1 dimensions
allows for a  controlled renormalization procedure because
 only two self-energies --
containing pure vacuum parts of Green functions -- diverge in the
ultraviolet regime. Thus the renormalization can be performed
equivalently to the standard zero temperature vacuum case. This is
different in 3+1 dimensions, where complete classes of
self-energies are divergent, which incorporate non-vacuum parts of
the Green functions as well. In the latter case more involved
renormalization schemes should be used analogous to those
constructed for self-consistent approximations at finite
temperature \cite{knollren1,knollren2,ren3}.

Our work is organized as follows: In Section 2 we will present the
relevant equations for the nonequilibrium dynamics in case of the
$\phi^4$-theory and provide the Kadanoff-Baym equations within the
three-loop approximation of the 2PI effective action. Furthermore,
we briefly review the derivation of generalized transport
equations and discuss different limits. Section 3 is devoted to
numerical studies on equilibration phenomena within the full
Kadanoff-Baym theory (cf. Ref. \cite{Juchem03}) in comparison to
the generalized transport equations in either Kadanoff-Baym (KB)
 form \cite{KB}
or Botermans-Malfliet (BM) approximation \cite{botmal}. We will also address the
differences in the solutions of the two different versions (KB and
BM) of the off-shell transport theory and explore the validity of
the gradient expansion employed. We close this work in Section IV
with a summary of our results and an outlook on future extensions.

\section{\label{sec:general_transport_derivation} Derivation of
Generalized Transport Equations}

In continuation of our previous study in Ref. \cite{Juchem03} we
investigate the out-of-equilibrium quantum dynamics of interacting
scalar fields by the example of the $\phi^4$-theory that allows to
test theoretical approximations in a variety of aspects
\cite{Peter,Peter2,berges1,berges2}. Its Lagrangian density is
given by $(x=(t,\vec{x}))$\\
\bea \label{lagrangian} {\cal L}(x) \; = \;
  \frac{1}{2} \, \partial_{\mu} \phi(x) \, \partial^{\mu} \phi(x)
\: - \: \frac{1}{2} \, m^2 \, \phi^2(x) \: - \: \frac{\lambda}{4
!} \, \phi^4(x) \; , \eea\\
where $m$ denotes the `bare' mass and $\lambda$ is the coupling
constant determining the interaction strength of the scalar
fields.

For the derivation and investigation of generalized transport
equations we first have to specify the necessary notations and the
explicit form of the Kadanoff-Baym equations. Though the details
of the derivation are given in Ref. \cite{Juchem03} we briefly
recall the main steps in order to keep the paper self-containt.

\subsection{\label{sec:theorykaba} The Kadanoff-Baym Equations}

A natural starting point for
nonequilibrium quantum theory is provided by the closed-time-path
(CTP) method. Here all quantities are given on a special real-time
contour with the time argument running from $-\infty$ to $+\infty$
on the chronological branch $(+)$ and returning from $+\infty$ to
$-\infty$ on the anti-chronological branch $(-)$. In particular the
path-ordered Green
functions are defined as $(x_1 = (t_1,\vec{x}_1),\; x_2 =
(t_2,\vec{x}_2))$\\[-0.3cm]
\bea \label{pathgreendef} G(x_1,x_2) & = & \langle \: T^p \, \{ \,
\phi(x_1) \: \phi(x_2) \, \} \, \rangle \\[0.4cm] & = &
\Theta_p(t_1-t_2) \; \langle \, \phi(x_1) \: \phi(x_2) \, \rangle
\;+\; \Theta_p(t_2-t_1) \; \langle \, \phi(x_2) \: \phi(x_1) \,
\rangle \, , \nn \eea\\[-0.3cm]
where the operator $T^p$ orders the field operators according to
the position of their arguments on the real-time path as
accomplished by the path step-functions $\Theta_p$ \\[-0.2cm]
\bea \Theta_p(t_1-t_2) \; = \; \left\{
\begin{array}{cl}
1 \, , \;\; & {\rm if\;\:} t_1 \;\: {\rm
is\;later\;on\;the\;contour\;as\;\:} t_2 \, , \\[0.4cm] 0 \, ,
\;\; & {\rm if\;\:} t_2 \;\: {\rm
is\;earlier\;on\;the\;contour\;as\;\:} t_1 \, .
\end{array}
\right. \eea\\[-0.2cm]
The expectation value in (\ref{pathgreendef}) is taken with
respect to some given density matrix $\rho_0$, which is constant
in time, while the operators in the Heisenberg picture contain the
whole information of the time dependence of the nonequilibrium
system.

Self-consistent equations of motion for these Green functions can
be obtained with help of the
two-particle-irreducible\footnote{n-particle-irreducible diagrams
remain connected when n arbitrary internal lines are cut.} (2PI)
effective action $\Gamma[G]$, which is given by the
double Legendre transform of the generating functional of
connected Green functions $W$ with respect to the local and
bilinear external sources \cite{Juchem03}.  The
functional $\Phi$, which is the sum of all closed
2PI diagrams built up by full propagators $G$ (and bare vertices)
\cite{knoll1,berges3,luttinger}, determines the resummed self-energy by
functional variation as\\[-0.2cm]
\bea \label{effactionsigma} \Sigma(x_1,x_2) \; = \; 2 i \,
\frac{\delta \Phi}{\delta G(x_2,x_1)} \: . \eea\\[-0.2cm]
In general, the self-energy $\Sigma$ contains a singular
contribution $\Sigma^{\delta}$ on the contour and a non-local
part, which can be expressed  by a sum over
$\Theta_p$-functions\\[-0.2cm]
\bea \label{pathselfdef} \Sigma(x_1,x_2) & = &
\Sigma^{\delta}(x_1) \; \delta_p^{(d+1)}(x_1\!-\!x_2) \\[0.4cm] &&
+ \; \Theta_p(t_1\!-\!t_2) \: \Sigma^{>}(x_1,x_2)      \; + \;
\Theta_p(t_2\!-\!t_1) \: \Sigma^{<}(x_1,x_2) \, . \phantom{a} \nn
\eea\\[-0.2cm]
Here the $\delta_p$-function on the closed-time-path is defined as
\\[-0.2cm]
\bea &&\delta_p^{(d+1)}(x_1-x_2) \;=\; \delta_p(t_1-t_2) \;\;
\delta^{(d)}(\vec{x}_1-\vec{x}_2) \, ,\\[0.4cm]
&&\delta_p(t_1-t_2) \;=\; \left\{
\begin{array}{cl}
\phantom{-} \delta(t_1-t_2) \, , \;\; & {\rm if\;both\;\:} t_1
\;\: {\rm and\;\:} t_2 \;\: {\rm are\;on\;the\;upper\;(+)\;branch}
\, , \nnl[0.2cm] - \delta(t_1-t_2) \, , \;\; & {\rm if\;both\;\:}
t_1 \;\: {\rm and\;\:} t_2 \;\: {\rm
are\;on\;the\;lower\;(-)\;branch} \, , \nnl[0.2cm] 0 \, , \;\; &
{\rm in\;other\;cases} \, ,
\end{array}
\right. \eea\\
in accordance with the contour integration while $d$ denotes the
spatial dimension of the problem. The functional $\Phi$ allows --
by restricting to a particular set of diagrams -- for the
construction of effective theories, which are conserving and
thermodynamically consistent \cite{knoll1,baym}. In our present
calculations we will take into account contributions up to the
three-loop order for the $\Phi$-functional (cf. Fig. 1 in Ref.
\cite{Juchem03}). Whereas the self-energies are defined by
(\ref{effactionsigma}) the evolution of the Green function is
determined by the stationarity condition $\delta \Gamma/\delta G =
0$ \cite{knoll6,Juchem03}.

For actual calculations it is advantageous to change to a
single-time representation for the Green functions and (the
non-local part of) the self-energies defined on the
closed-time-path. In line with the position of the coordinates on
the contour there exist four different two-point functions
\\[-0.2cm]
\bea \label{green_def} i \, G^{c}(x_1,x_2) & = & i \,
G^{++}(x_1,x_2) \;\; = \;\; i \, G(x_1^+,x_2^+)  \;\; = \;\;
\langle \, T^c \, \{ \, \phi(x_1) \: \phi(x_2) \, \} \, \rangle \,
, \phantom{aaa} \\[0.3cm] i \, G^{<}(x_1,x_2) & = & i \,
G^{+-}(x_1,x_2) \;\; = \;\; i \, G(x_1^+,x_2^-)  \;\; = \;\;\;
\;\;\;\: \langle \, \{ \, \phi(x_2) \: \phi(x_1) \, \} \, \rangle
\, , \nnl[0.3cm] i \, G^{>}(x_1,x_2) & = & i \, G^{-+}(x_1,x_2)
\;\; = \;\; i \, G(x_1^-,x_2^+) \;\; = \;\;\: \;\;\;\: \langle \,
\{ \, \phi(x_1) \: \phi(x_2) \, \} \, \rangle \, , \nnl[0.3cm] i
\, G^{a}(x_1,x_2) & = & i \, G^{--}(x_1,x_2) \;\; = \;\; i \,
G(x_1^-,x_2^-) \;\; = \;\; \langle \, T^a \, \{ \, \phi(x_1) \:
\phi(x_2) \, \} \, \rangle \, . \nn \eea\\[-0.2cm]
Here $T^c \, (T^a)$ represent the (anti-)time-ordering operators
in case of both arguments lying on the (anti-)chronological branch
of the real-time contour. These four functions are not independent
of each other. In particular the non-continuous functions $G^c$
and $G^a$ are built up by the Wightman functions $G^>$ and $G^<$
and the usual $\Theta$-functions in the time coordinates as
\\[-0.2cm]
\bea G^c(x_1,x_2) & = &
%\langle \: T^c \, \{ \, \phi(x_1) \: \phi(x_2) \, \} \, \rangle \;=\;
         \Theta(t_1-t_2) \; G^{>}(x_1,x_2) \:+\:
         \Theta(t_2-t_1) \; G^{<}(x_1,x_2) \, , \\[0.4cm]
G^a(x_1,x_2) & = &
%\langle \: T^a \, \{ \, \phi(x_1) \: \phi(x_2) \, \} \, \rangle \;=\;
         \Theta(t_2-t_1) \; G^{>}(x_1,x_2) \:+\:
         \Theta(t_1-t_2) \; G^{<}(x_1,x_2) \, .
\nn \eea\\[-0.2cm]
By fixing the time arguments of the Green functions on different
branches of the contour and resolving the time structure of the
path-ordered quantities  we obtain the Kadanoff-Baym equations for the time
evolution of the Wightman functions
\cite{danmrow,Juchem03,berges1}:\\[-0.2cm]
\bea \label{kabaeqcs} - \left[
\partial_{\mu}^{x_1} \partial_{x_1}^{\mu} \!+ m^2
\right] \, G^{\gtrless}(x_1,x_2) & = & \Sigma^{\delta}(x_1) \;
G^{\gtrless}(x_1,x_2) \\[0.5cm] && \!\!\!\!\!\!\!\!\!\!
\!\!\!\!\!\!\!\!\!\! + \int_{t_0}^{t_1} \!\!\!\!\! dz_0 \int
\!\!d^{d}\!z \;\; \left[\,\Sigma^{>}(x_1,z) - \Sigma^{<}(x_1,z)
\,\right] \: G^{\gtrless}(z,x_2) \nnl[0.4cm] &&
\!\!\!\!\!\!\!\!\!\! \!\!\!\!\!\!\!\!\!\! - \int_{t_0}^{t_2}
\!\!\!\!\! dz_0 \int \!\!d^{d}\!z \;\;\, \Sigma^{\gtrless}(x_1,z)
\: \left[\,G^{>}(z,x_2) - G^{<}(z,x_2) \,\right] \, , \nnl[0.7cm]
- \left[
\partial_{\mu}^{x_2} \partial_{x_2}^{\mu} \!+ m^2
\right] \, G^{\gtrless}(x_1,x_2) & = & \Sigma^{\delta}(x_2) \;
G^{\gtrless}(x_1,x_2) \nnl[0.5cm] && \!\!\!\!\!\!\!\!\!\!
\!\!\!\!\!\!\!\!\!\! + \int_{t_0}^{t_1} \!\!\!\!\! dz_0 \int
\!\!d^{d}\!z \;\; \left[\,G^{>}(x_1,z) - G^{<}(x_1,z) \,\right] \:
\Sigma^{\gtrless}(z,x_2) \nnl[0.4cm] && \!\!\!\!\!\!\!\!\!\!
\!\!\!\!\!\!\!\!\!\! - \int_{t_0}^{t_2} \!\!\!\!\! dz_0 \int
\!\!d^{d}\!z \;\;\, G^{\gtrless}(x_1,z) \:
\left[\,\Sigma^{>}(z,x_2) - \Sigma^{<}(z,x_2) \,\right] \, . \nn
\eea\\
Within the three-loop approximation for the 2PI effective action
$\Phi$ we get two
different self-energies: In leading order of the coupling constant
only the local tadpole diagram  contributes and leads to the
generation of an effective mass for the field quanta. This
self-energy (in coordinate space) is given by \cite{Juchem03} \\[-0.2cm]
\bea \label{tadpole_cs} \Sigma^{\delta}(x) \; = \;
\frac{\lambda}{2} \; i \: G^{<}(x,x) \: . \eea\\[-0.2cm]
In next order in the coupling constant (i.e. $\lambda^2$) the
non-local sunset self-energy (cf. Fig. 2 in \cite{Juchem03})
 enters the time evolution as\\[-0.2cm]
\bea \label{sunset_cs} \Sigma^{\gtrless}(x_1,x_2) \; = \; -
\frac{\lambda^2}{6} \; G^{\gtrless}(x_1,x_2) \;
G^{\gtrless}(x_1,x_2) \; G^{\lessgtr}(x_2,x_1) \; = \; -
\frac{\lambda^2}{6} \; \left[ \, G^{\gtrless}(x_1,x_2) \,
\right]^3 \!\!\! . \phantom{a} \eea\\[0.2cm]
Thus the Kadanoff-Baym equation (\ref{kabaeqcs}) in our case
includes the influence of a mean-field on the particle propagation
-- generated by the tadpole diagram -- as well as  scattering
processes as inherent in the sunset diagram.

The Kadanoff-Baym equation (\ref{kabaeqcs}) describes the full
quantum nonequilibrium time evolution on the two-point level for a
system prepared at an initial time $t_0$, i.e. when higher order
correlations are discarded. The causal structure of this initial
value problem is obvious since the time integrations are performed
over the past up to the actual time $t_1$ (or $t_2$, respectively)
and do not extend to the future.

Furthermore, also linear combinations of the Green functions in
single-time representation are of interest and will be exploited
for the spectral properties of the system later on. The retarded
Green function $G^R$ and the advanced Green function $G^A$ are
given as\\[-0.2cm]
\bea \label{defret} G^{R}(x_1,x_2) & = & \phantom{-} \, \Theta(t_1
- t_2) \; \left[ \, G^{>}(x_1,x_2) - G^{<}(x_1,x_2) \, \right]
\\[0.3cm]
& = & \phantom{-} \, \Theta(t_1 - t_2) \; \langle \, \left[ \,
\phi(x_1) \, , \, \phi(x_2) \, \right]_{-} \, \rangle \nnl[0.3cm]
& = & \phantom{-} \, G^{c}(x_1,x_2) \:-\: G^{<}(x_1,x_2)
         \;\: = \;\: G^{>}(x_1,x_2) \:-\: G^{a}(x_1,x_2) \, ,
\nnl[0.5cm] \label{defadv} G^{A}(x_1,x_2) & = & - \, \Theta(t_2 -
t_1) \; \left[ \, G^{>}(x_1,x_2) - G^{<}(x_1,x_2) \, \right]
\\[0.3cm]
& = & - \, \Theta(t_2 - t_1) \; \langle \, \left[ \, \phi(x_1) \,
, \, \phi(x_2) \, \right]_{-} \, \rangle \nnl[0.3cm] & = &
\phantom{-} \, G^{c}(x_1,x_2) \:-\: G^{>}(x_1,x_2) \;\: = \;\:
G^{<}(x_1,x_2) \:-\: G^{a}(x_1,x_2) \, . \nn \eea\\[-0.2cm]
These Green functions contain exclusively spectral, but no
statistical information of the system. Their time evolution is
determined by  Dyson-Schwinger equations  and given by (cf.
Ref. \cite{Juchem03})\\[-0.2cm]
\bea \label{dseqretcs} && - \left[
\partial_{\mu}^{x_1} \partial_{x_1}^{\mu} \!+ m^2 + \Sigma^{\delta}(x_1)
\right] \, G^{R}(x_1,x_2) \;=\; \\[0.3cm] && \qquad \qquad \qquad
\qquad \qquad \delta^{(d+1)}(x_1\!-\!x_2) \:+\: \int \!d^{d+1}\!z
\;\; \Sigma^{R}(x_1,z) \;\; G^{R}(z,x_2) \, , \nnl[0.6cm]
\label{dseqadvcs} && - \left[
\partial_{\mu}^{x_1} \partial_{x_1}^{\mu} \!+ m^2 + \Sigma^{\delta}(x_1)
\right] \, G^{A}(x_1,x_2) \:=\:\\[0.3cm] && \qquad \qquad \qquad
\qquad \qquad \delta^{(d+1)}(x_1\!-\!x_2) \:+\: \int \!d^{d+1}\!z
\;\; \Sigma^{A}(x_1,z) \;\; G^{A}(z,x_2) \, , \nn \eea\\[-0.2cm]
where the retarded and advanced self-energies $\Sigma^{R}$,
$\Sigma^{A}$ are defined via $\Sigma^{>}$, $\Sigma^{<}$ similar to
the Green functions (\ref{defret}) and (\ref{defadv}). Thus the
retarded (advanced) Green functions are determined by retarded
(advanced) quantities, only. \\

\subsection{Generalized transport equations}
The derivation of generalized transport equations has been
presented before in Refs.
\cite{caju1,KB,GL98a,botmal,HHab,knoll3,knoll6,Leupold}. We here briefly repeat
the derivation in order to introduce the necessary notations and
to work out the differences in the Kadanoff-Baym \cite{KB} and
Botermans-Malfliet \cite{botmal} scheme that will be investigated
in Section 3.

We start by rewriting the
Kadanoff-Baym equation for the Wightman functions in coordinate space
($ x_1\!=\!(t_1,\vec{x}_1), x_2\!=\!(t_2,\vec{x}_2) $)
(\ref{kabaeqcs}) as\\
\bea
\label{eq:kbcs1}
[ \, \partial^{\mu}_{x_1} \partial_{\mu}^{x_1} + m^2 +
\Sigma^{\delta}(x_1) \, ] \;\:
i G^{\gtrless}(x_1,x_2)
\; = \; i\,I_1^{\gtrless}(x_1,x_2) \, .
\eea\\
The collision terms on the r.h.s. of (\ref{eq:kbcs1}) are given in
$D = d+1$ space-time dimensions
by convolution integrals over coordinate space self-energies and
Green functions: \\
\bea
\label{eq:i1cs}
I_1^{\gtrless}(x_1,x_2)
\: = \:
& - & \!\!
\int_{t_0}^{t_1} \!\!\! d^{D}\!z \; \;
\left[ \,
\Sigma^{>}(x_1,z) -
\Sigma^{<}(x_1,z)
\, \right] \;\;
G^{\gtrless}(z,x_2) \\[0.5cm]
& + & \!\!
\int_{t_0}^{t_2} \!\!\! d^{D}\!z \; \;
\Sigma^{\gtrless}(x_1,z) \; \;
\left[ \,
G^{>}(z,x_2) -
G^{<}(z,x_2)
\, \right] \, .
\nn
\eea\\
In the general case of an arbitrary (scalar) quantum field theory
$\Sigma^{\delta}$ is the local (non-dissipative) part of the path
self-energy while $\Sigma^{\gtrless}$ resemble the non-local
collisional self-energy contributions. In the representation
(\ref{eq:i1cs}) the integration boundaries are exclusively given
for the time coordinates since the integration over
 the spatial coordinates extends over
the whole spatial volume from $- \infty$ to $+ \infty$.

Since transport theories   are formulated in phase-space one
changes  to the Wigner representation via Fourier transformation
with respect to the rapidly varying ('intrinsic') relative
coordinate $\Delta x = x_1 - x_2$ and treats the system evolution
in terms of the ('macroscopic') mean space-time coordinate $x =
(x_1 + x_2)/2$ and the four-momentum $p = (p_0,\vec{p})$. The
functions in Wigner space are obtained as \\
\bea
\label{eq:wignertrafo}
\bar{F}(p,x) \; = \;
\int_{-\infty}^{\infty} \!\!\! d^{D}\!\Delta x \;\;
\; e^{+i\:\Delta x_{\mu}\:p^{\mu}} \;\;
F(x_1=x+\Delta x/2,\,x_2=x-\Delta x/2) \, .
\eea\\
For the formulation of transport theory in the Wigner representation
we have to focus not only on the transformation properties of
ordinary two-point functions as given in (\ref{eq:wignertrafo}),
but also of convolution integrals as appearing in Eq. (\ref{eq:i1cs}).
A convolution integral in $D$ dimensions (for arbitrary
functions $F, G$), \\
\bea
\label{eq:conv}
H(x_1,x_2) \; = \;
\int_{-\infty}^{\infty} \!\!\! d^{D}\!z \;\;
F(x_1,z) \;\; G(z,x_2)
\eea\\
transforms as \\
\bea
\bar{H}(p,x)
& = &
\int_{-\infty}^{\infty} \!\!\! d^{D}\!\Delta x \;\;
\; e^{+i\:\Delta x_{\mu}\:p^{\mu}} \;\;
H(x_1,x_2)
\\[0.4cm]
& = &
\int_{-\infty}^{\infty} \!\!\! d^{D}\!\Delta x \;\;
\; e^{+i\:\Delta x_{\mu}\:p^{\mu}} \;\;
\int_{-\infty}^{\infty} \!\!\! d^{D}\!z \;\;
F(x_1,z) \;\; G(z,x_2)
\nnl[0.6cm]
& = &
\left.
\; e^{+i\,\frac{1}{2} \,
(\partial_{p^{\phantom{\prime}}}^{\mu} \!\! \cdot \:
 \partial^{x^{\prime}}_{\mu}
 \: - \;
 \partial_{x^{\phantom{\prime}}}^{\mu} \!\! \cdot \:
 \partial^{p^{\prime}}_{\mu} ) } \;\:
\left[ \;
\bar{F}(p,x) \;\;
\bar{G}(p^{\prime},x^{\prime})
\; \right] \right|_{x^{\prime} = x,\: p^{\prime} = p} \; .
\nn
\eea\\
In accordance with the standard assumption of transport theory we
assume that all functions only smoothly evolve in the
mean space-time coordinates and thus restrict to first order derivatives.
All terms proportional to second or higher order
derivatives in the mean space-time coordinates (also mixed ones)
will be dropped.
Thus the Wigner transformed convolution integrals (\ref{eq:conv}) are given
in {\it first order gradient approximation} by, \\
\bea
\label{eq:firstordergrad}
\bar{H}(p,x)
\; = \;
\bar{F}(p,x) \;\;
\bar{G}(p,x)
\; + \;
i \, \frac{1}{2} \:
\{ \, \bar{F}(p,x) \, , \, \bar{G}(p,x) \, \}
\; + \; {\cal O}(\partial^2_x) \, ,
\eea\\
using the relativistic generalization of the Poisson bracket \\
\bea
\label{eq:def_poisson}
\{ \, \bar{F}(p,x) \, , \, \bar{G}(p,x) \, \}
\; = \;
\partial^{p}_{\mu} \, \bar{F}(p,x) \cdot
\partial_{x}^{\mu} \, \bar{G}(p,x) \; - \;
\partial_{x}^{\mu} \, \bar{F}(p,x) \cdot
\partial^{p}_{\mu} \, \bar{G}(p,x) \; .
\eea\\

In order to obtain the dynamics for the spectral functions within
the approximate scheme we start with the Dyson-Schwinger equations
for the retarded and advanced Green functions in coordinate space
(\ref{dseqretcs},\ref{dseqadvcs}). -- We note that the convolution
integrals in (\ref{dseqretcs}) and (\ref{dseqadvcs}) extend over
the whole space and time range in contrast to the equations of
motion for the Wightman functions given in (\ref{eq:kbcs1}) and
(\ref{eq:i1cs}). -- The further procedure consists in the
following steps: First we
\\
i) transform the above equations into the Wigner representation
and apply the first order gradient approximation.
In this limit the convolution integrals yield
the product terms and the general Poisson bracket of
the self-energies and the Green functions $\{\, \Sigma^{R/A}, G^{R/A} \,\}$.
We, furtheron, represent both equations in terms of
real quantities by the decomposition of the retarded
and advanced Green functions and self-energies as \\
\bea
\begin{array}{ccccccc}
\bar{G}^{R/A}
&\!\!=\!\!& Re\,\bar{G}^{R} \,\pm\, i\,Im\,\bar{G}^{R}
&\!\!=\!\!& Re\,\bar{G}^{R} \,\mp\, i\,\bar{A} / 2\; ,
\phantom{aaaaaa}
\bar{A} &\!\!=\!\!& \mp \, 2 \, Im\,\bar{G}^{R/A} \, ,
\\[0.7cm]
\bar{\Sigma}^{R/A}
&\!\!=\!\!& Re\,\bar{\Sigma}^{R} \,\pm\, i\,Im\,\bar{\Sigma}^{R}
&\!\!=\!\!& Re\,\bar{\Sigma}^{R} \,\mp\, i\,\bar{\Gamma} / 2 \; ,
\phantom{aaaaaa}
\bar{\Gamma} &\!\!=\!\!& \mp \, 2\, Im\,\bar{\Sigma}^{R/A} \, .
\end{array}
\phantom{aa}
\eea\\
We find that in Wigner space the real parts of the retarded and
advanced Green functions and self-energies are equal, while the
imaginary parts have opposite sign and are proportional to the
spectral function $\bar{A}$ and the width $\bar{\Gamma}$,
respectively. The next step consists in \\ ii) the separation of
the real part and the imaginary part of the two equations for the
retarded and advanced Green functions, that have to be fulfilled
independently. Thus we obtain four real-valued equations for the
self-consistent retarded and advanced Green functions. In the last
step \\ iii) we get simple relations by linear combination of
these equations, i.e. by adding/subtrac\-ting the relevant
equations.

This finally leads to two algebraic relations for
the spectral function $\bar{A}$ and the real part of the retarded
Green function $Re\,\bar{G}^{R}$ in terms of the width $\bar{\Gamma}$
and the real part of the retarded self-energy $Re\,\bar{\Sigma}^{R}$
as \cite{caju1,caju2}: \\
\bea
\label{eq:specrel1}
[ \, p_0^2 - \vec{p}^{\,2} - m^2 -
\bar{\Sigma}^{\delta} + Re\,\bar{\Sigma}^{R} \, ] \; Re\,\bar{G}^{R}
& = &
1 \: + \: \frac{1}{4} \: \bar{\Gamma} \; \bar{A} \, ,
\\[0.6cm]
\label{eq:specrel2}
[ \, p_0^2 - \vec{p}^{\,2} - m^2 -
\bar{\Sigma}^{\delta} + Re\,\bar{\Sigma}^{R} \, ] \; \bar{A}
& = &
\bar{\Gamma} \; Re\,\bar{G}^{R} \, .
\eea\\
Note that all terms with first order gradients have disappeared in
(\ref{eq:specrel1}) and (\ref{eq:specrel2}). A first consequence
of (\ref{eq:specrel2}) is a direct relation between the real and
the imaginary parts of the retarded/advanced Green function, which
reads (for $\bar{\Gamma} \neq 0$): \\
\bea \label{ins1}
 Re\,\bar{G}^{R} \; = \; \frac{p_0^2 -
\vec{p}^{\,2} - m^2 - \bar{\Sigma}^{\delta} - Re\,\bar{\Sigma}^{R}
}{\bar{\Gamma}} \; \bar{A} \; . \eea\\
Inserting (\ref{ins1}) in (\ref{eq:specrel1})  we end up with the following result for the
spectral function and the real part of the retarded Green function
\\
\bea
\label{eq:specorder0}
\bar{A} \; = \;
\frac{\bar{\Gamma}}{[ \, p_0^2 - \vec{p}^{\,2} - m^2
- \bar{\Sigma}^{\delta} - Re\,\bar{\Sigma}^{R} \, ]^2 + \bar{\Gamma}^2/4}
& = &
\frac{\bar{\Gamma}}{\bar{M}^2 + \bar{\Gamma}^2/4} \, , \phantom{aaa}
\\[0.6cm]
\label{eq:regretorder0}
Re\,\bar{G}^{R} \; = \;
\frac{[ \, p_0^2 - \vec{p}^{\,2} - m^2
- \bar{\Sigma}^{\delta} - Re\,\bar{\Sigma}^{R} \, ]}
{[ \, p_0^2 - \vec{p}^{\,2} - m^2
- \bar{\Sigma}^{\delta} - Re\,\bar{\Sigma}^{R} \, ]^2 + \bar{\Gamma}^2/4}
& = &
\frac{\bar{M}}{\bar{M}^2 + \bar{\Gamma}^2/4} \, ,
\eea\\
where we have introduced the mass-function $\bar{M}$ in Wigner space: \\
\bea
\label{eq:massfunction}
\bar{M}(p,x) & = & p_0^2 - \vec{p}^{\,2} - m^2
- \bar{\Sigma}^{\delta}(x) - Re\,\bar{\Sigma}^{R}(p,x) \; .
\eea\\
The  spectral function (\ref{eq:specorder0}) shows a typical Breit-Wigner
shape with energy- and momentum-dependent self-energy terms.
Although the above equations are purely algebraic solutions and
contain no derivative terms, they are valid up to the first order in the
gradients.

In addition, subtraction of the real parts and adding up the imaginary
parts lead to the time evolution equations \\
\bea
\label{eq:specorder1}
p^{\mu} \, \partial_{\mu}^x \, \bar{A}
& = &
\frac{1}{2} \,
\{ \, \bar{\Sigma}^{\delta} + Re\,\bar{\Sigma}^{R} \, ,
   \, \bar{A} \, \}
\: + \: \frac{1}{2} \,
\{ \, \bar{\Gamma} \, , \, Re\,\bar{G}^{R} \, \} \, ,
\\[0.6cm]
p^{\mu} \, \partial_{\mu}^x \, Re\,\bar{G}^{R}
& = &
\label{eq:regretorder1}
\frac{1}{2} \,
\{ \, \bar{\Sigma}^{\delta} + Re\,\bar{\Sigma}^{R} \, ,
   \, Re\,\bar{G}^{R} \, \}
\: - \: \frac{1}{8} \,
\{ \, \bar{\Gamma} \, , \, \bar{A} \, \} \, .
\eea\\
The Poisson bracket containing the mass-function $\bar{M}$ leads to
the well-known drift operator $p^{\mu}\,\partial^{x}_{\mu}\,\bar{F}$
(for an arbitrary function $\bar{F}$), i.e. \\
\bea
\label{eq:mass_poisson}
\{\, \bar{M} \, , \, \bar{F} \,\}
& = &
\{\, p_0^2 - \vec{p}^{\,2} - m^2 -
\bar{\Sigma}^{\delta} - Re\,\bar{\Sigma}^{R}
\, , \, \bar{F} \,\} \\[0.6cm]
& = &
2 \, p^{\mu} \, \partial_{\mu}^x \: \bar{F} \: - \:
\{\,\bar{\Sigma}^{\delta} + Re\,\bar{\Sigma}^{R} \, , \, \bar{F} \,\} \; ,
\eea\\
such that the first order equations (\ref{eq:specorder1}) and
(\ref{eq:regretorder1}) can be written in a more
comprehensive form as \\
\bea
\label{eq:specorder1final}
\{ \, \bar{M} \, , \, \bar{A} \, \}
& = &
\{ \, \bar{\Gamma} \, , \, Re\,\bar{G}^{R} \, \} \, ,
\\[0.6cm]
\{ \, \bar{M} \, , \, Re\,\bar{G}^{R} \, \}
& = &
\label{eq:regretorder1final}
- \, \frac{1}{4} \,
\{ \, \bar{\Gamma} \, , \, \bar{A} \, \} \, .
\eea\\
When inserting (\ref{eq:specorder0}) and (\ref{eq:regretorder0})
we find that these first order time evolution equations are {\em
solved} by the algebraic expressions. In this case the following
relations hold: \\
\bea
\{\, \bar{M} \, , \, \bar{A} \,\}
\; = \;
\{\, \bar{\Gamma} \, , \, Re\,\bar{G}^{R} \,\}
& = &
\{\, \bar{M} \, , \, \bar{\Gamma} \,\} \;\;
\frac{\bar{M}^2 - \bar{\Gamma}^2/4}
{[\, \bar{M}^2 + \bar{\Gamma}^2/4 \,]^2} \, ,
\\[0.6cm]
\{\, \bar{M} \, , \, Re\,\bar{G}^{R} \,\}
\; = \;
- \, \frac{1}{4} \, \{\, \bar{\Gamma} \, , \, \bar{A} \,\}
& = &
\{\, \bar{M} \, , \, \bar{\Gamma} \,\} \;\;
\frac{ \bar{M} \, \bar{\Gamma} / 2 }
{[\, \bar{M}^2 + \bar{\Gamma}^2/4 \,]^2} \, .
\eea\\
Thus we have derived the proper structure of the
spectral function (\ref{eq:specorder0})
within the first-order gradient (or  semiclassical) approximation.
Together with the explicit form for the real part of the retarded Green
function (\ref{eq:regretorder0}) we now have fixed the dynamics
of the spectral properties, which is consistent up the first order
in the gradients.

As a next step we rewrite the memory terms in the collision
integrals (\ref{eq:i1cs})
such that the time integrations extend from $- \infty$ to
$+ \infty$.
In this respect we consider the initial time
$t_0 = - \infty$ whereas the upper time boundaries $t_1, t_2$ are
taken into account by $\Theta$-functions, i.e. \\
\bea
\label{eq:i1csnew}
I_1^{\gtrless}(x_1,x_2)
\: = \:
& - & \!\!
\int_{-\infty}^{\infty} \!\!\! d^{D}x^{\prime} \;\;\;
\Theta(t_1-t^{\prime}) \:
\left[ \,
\Sigma^{>}(x_1,x^{\prime}) -
\Sigma^{<}(x_1,x^{\prime})
\, \right] \;\;
G^{\gtrless}(x^{\prime},x_2)
\nnl[0.5cm]
& + & \!\!
\int_{-\infty}^{\infty} \!\!\! d^{D}x^{\prime} \;\;\;
\Sigma^{\gtrless}(x_1,x^{\prime}) \; \;
\Theta(t_2-t^{\prime}) \:
\left[ \,
G^{>}(x^{\prime},x_2) -
G^{<}(x^{\prime},x_2)
\, \right]
\nnl[1.0cm]
\: = \:
& - & \!\!
\int_{-\infty}^{\infty} \!\!\! d^{D}x^{\prime} \;\;\;\;
\Sigma^{R}(x_1,x^{\prime}) \;
G^{\gtrless}(x^{\prime},x_2) \: + \:
\Sigma^{\gtrless}(x_1,x^{\prime}) \;
G^{A}(x^{\prime},x_2) \; . \phantom{aaaa}
\eea\\
We now perform the analogous steps as invoked before for the
retarded and advanced Dyson-Schwinger equations. We start with a
first order gradient expansion of the Wigner transformed
Kadanoff-Baym equation using (\ref{eq:i1csnew}) for the memory
integrals. Again we separate the real and the imaginary parts in
the resulting equation, which have to be satisfied independently.
At the end of this procedure we obtain a generalized transport
equation \cite{caju1,caju2,KB,GL98a,botmal,knoll3,knoll6,Leupold}: \\
\bea
\label{eq:general_transport}
\underbrace{
\phantom{\frac{1}{1}} \!\!\!
2\,p^{\mu}\:\partial^{x}_{\!\mu} \: i\bar{G}^{\gtrless}
\, - \,
\{ \, \bar{\Sigma}^{\delta} \!+\! Re\,\bar{\Sigma}^{R} ,
   \, i \bar{G}^{\gtrless} \, \} }
\, - \,
\{ \, i\bar{\Sigma}^{\gtrless} \, , \, Re\,\bar{G}^{R} \, \}
& = &
i\bar{\Sigma}^{<} \; i\bar{G}^{>}
\, - \,
i\bar{\Sigma}^{>} \; i\bar{G}^{<}
\nnl[0.4cm]
\{ \, \bar{M} \, , \, i \bar{G}^{\gtrless} \, \}  \;\; \qquad \qquad
\, - \,
\{ \, i\bar{\Sigma}^{\gtrless} \, , \, Re\,\bar{G}^{R} \, \}
& = &
i\bar{\Sigma}^{<} \; i\bar{G}^{>}
\, - \,
i\bar{\Sigma}^{>} \; i\bar{G}^{<} \phantom{aaaaaaa}
\eea\\
as well as a generalized mass-shell equation \\
\bea
\label{eq:general_mass}
\underbrace{
\phantom{\frac{1}{1}} \!\!\!\!
[ \, p^2 - m^2 - \bar{\Sigma}^{\delta} - Re\,\bar{\Sigma}^{R} \, ]}_{\bar{M}}
\;\,
i \bar{G}^{\gtrless}
\: = \:
i\bar{\Sigma}^{\gtrless} \; Re\,\bar{G}^{R}
\, + \, \frac{1}{4} \, \{ \,
i\bar{\Sigma}^{>} , \, i\bar{G}^{<} \, \}
\, - \, \frac{1}{4} \, \{ \,
i\bar{\Sigma}^{<} , \, i\bar{G}^{>} \, \}  \phantom{aaa}
\eea\\
with the mass-function $\bar{M}$ specified in (\ref{eq:massfunction}).
Since the Green function $G^{\gtrless}(x_1,x_2)$ consists of an
antisymmetric real part and a symmetric imaginary part with respect
to the relative coordinate $x_1-x_2$, the Wigner transform of this
function is purely imaginary.
It is thus convenient to represent the Wightman functions in Wigner
space furtheron by the real-valued quantities $i \bar{G}^{\gtrless}(p,x)$.
Since the collisional self-energies obey the same symmetry relations
in coordinate space and in phase-space, they will be kept also as
$i \bar{\Sigma}^{\gtrless}(p,x)$ furtheron.

In the transport equation (\ref{eq:general_transport}) one
recognizes on the l.h.s. the drift term
$p^{\mu}\:\partial^{x}_{\!\mu} \: i\bar{G}^{\gtrless}$, as well as
the Vlasov term with the local self-energy $\bar{\Sigma}^{\delta}$
and the real part of the retarded self-energy
$Re\,\bar{\Sigma}^{R}$. On the other hand the r.h.s. represents
the collision term with its typical `gain and loss' structure. The
loss term $i\bar{\Sigma}^{>} \; i\bar{G}^{<}$ (proportional to the
Green function itself) describes the scattering out of a
respective phase-space cell whereas the gain term
$i\bar{\Sigma}^{<} \; i\bar{G}^{>}$ takes into account scatterings
into the actual cell. The last term on the l.h.s. $\{\,
i\bar{\Sigma}^{\gtrless} , Re\,\bar{G}^{R} \, \}$ is very {\em
peculiar} since it does not contain directly the distribution
function $i\bar{G}^{<}$. This second Poisson bracket vanishes in
the quasiparticle approximation and thus does not appear in the
on-shell Boltzmann limit. As demonstrated in detail in Refs.
\cite{caju1,caju2,KB,knoll3,knoll6,Leupold}
 the second Poisson bracket $\{ \,
i\bar{\Sigma}^{\gtrless} , Re\,\bar{G}^{R} \, \}$ governs the
evolution of the off-shell dynamics for nonequilibrium systems.

Although the generalized transport equation
(\ref{eq:general_transport}) and the generalized mass-shell
equation (\ref{eq:general_mass}) have been derived from the same
Kadanoff-Baym equation in a first order gradient expansion, both
equations are not exactly equivalent \cite{caju1,botmal,knoll6}.
Instead, they deviate from each other by contributions of higher
gradient order (see below and Refs. \cite{caju1,botmal}). This
raises the question: {\it which one of these two equations has be
considered to higher priority?} The question is answered in
practical applications by the prescription of solving the
generalized transport equation (\ref{eq:general_transport}) for
$i\bar{G}^{<}$ in order to study the dynamics of the
nonequilibrium system in phase-space. Since the dynamical
evolution of the spectral properties is taken into account by the
equations derived in first order gradient expansion from the
retarded and advanced Dyson-Schwinger equations, one can neglect
the generalized mass-shell equation (\ref{eq:general_mass}). Thus
for our actual numerical studies in Section 3 we will  use the
generalized transport equation (\ref{eq:general_transport})
supported by the algebraic relations (\ref{eq:specorder0}) and
(\ref{eq:regretorder0}).

Furthermore, one recognizes by subtraction of the $i\bar{G}^{>}$ and
$i\bar{G}^{<}$ mass-shell and transport equations, that the dynamics
of the spectral function $\bar{A} = i\bar{G}^{>} - i\bar{G}^{<}$ is
determined in the same way as derived from the retarded and advanced
Dyson-Schwinger equations (\ref{eq:specorder0}) and
(\ref{eq:specorder1final}).
The inconsistency between the two equations
(\ref{eq:general_transport}) and (\ref{eq:general_mass})
vanishes  since the differences are
contained in the collisional contributions on the r.h.s.
of (\ref{eq:general_transport}).

In order to evaluate the
$\{\, i\bar{\Sigma}^{<} , Re\,\bar{G}^{R} \,\}$-term
on the l.h.s. of (\ref{eq:general_transport}) it is useful to
introduce distribution functions
for the Green functions and self-energies as \\
\bea
\label{eq:ansatz_dist}
i\bar{G}^{<}(p,x) \; = \;
\bar{N}(p,x) \; \bar{A}(p,x) \: ,
\quad &&
i\bar{G}^{>}(p,x) \; = \;
[ \, 1 \, + \, \bar{N}(p,x) \, ] \; \bar{A}(p,x) \: ,
\\[0.6cm]
i\bar{\Sigma}^{<}(p,x) \; = \;
\bar{N}^{\Sigma}(p,x) \; \bar{\Gamma}(p,x) \: ,
\quad &&
i\bar{\Sigma}^{>}(p,x) \; = \;
[ \, 1 \, + \, \bar{N}^{\Sigma}(p,x) \, ] \; \bar{\Gamma}(p,x) \: .
\phantom{aaaa}
\eea\\
In equilibrium the distribution function with respect to the
Green functions $\bar{N}$ and the self-energies $\bar{N}^{\Sigma}$
are given as Bose functions in the energy $p_0$ at given temperature and
thus are equal.
Following the argumentation of Botermans and Malfliet \cite{botmal}
the distribution functions $\bar{N}$ and $\bar{N}^{\Sigma}$ in
(\ref{eq:ansatz_dist})
should be identical within the second term of the l.h.s. of
(\ref{eq:general_transport}) within a consistent first order gradient
expansion. In order to demonstrate their argument we write
\\
\bea
i\bar{\Sigma}^{<}
\; = \; \bar{\Gamma} \; \bar{N}^{\Sigma}
\; = \; \bar{\Gamma} \; \bar{N} \; + \; \bar{K} \: .
\eea\\
The `correction' term \\
\bea
\bar{K} \; = \; \bar{\Gamma} \;
( \, \bar{N}^{\Sigma} \; - \; \bar{N} \, )
\; = \;
( \, i\bar{\Sigma}^{<} \; i\bar{G}^{>} \; - \;
     i\bar{\Sigma}^{>} \; i\bar{G}^{<} \,) \; \bar{A}^{-1} \: ,
\eea\\
is proportional to the collision term  of the generalized
transport equation (\ref{eq:general_transport}), which itself is
of first order in the gradients. Thus, whenever a distribution
function $\bar{N}^{\Sigma}$ appears within a Poisson bracket, the
difference term $(\bar{N}^{\Sigma} - \bar{N}$) becomes of second
order in the gradients and should be omitted for consistency. As a
consequence $\bar{N}^{\Sigma}$ can be replaced by $\bar{N}$ and
thus the self-energy $\bar{\Sigma}^{<}$ by $\bar{G}^< \cdot
\bar{\Gamma} / \bar{A}$ in the Poisson bracket term $\{
\bar{\Sigma}^{<} , Re\,\bar{G}^{R} \}$. The generalized transport
equation (\ref{eq:general_transport}) then can be written in
short-hand notation \\
\bea
\label{eq:general_transport_bm}
\frac{1}{2} \, \bar{A} \; \bar{\Gamma} \;
\left[ \, \{ \, \bar{M} \, , \, \, i\bar{G}^{<} \, \} \: - \:
\frac{1}{\bar{\Gamma}} \; \{ \, \bar{\Gamma} \, , \,
\bar{M} \cdot i\bar{G}^{<} \, \} \, \right]
\; = \;
i\bar{\Sigma}^{<} \; i\bar{G}^{>}
\, - \,
i\bar{\Sigma}^{>} \; i\bar{G}^{<} \;\;\;\;\;\;
\eea\\
with the mass-function $\bar{M}$ (\ref{eq:massfunction}). The
transport equation (\ref{eq:general_transport_bm}) within the
Botermans-Malfliet (BM) form resolves the discrepancy between the
generalized mass-shell equation (\ref{eq:general_mass}) and the
generalized transport equation in its original Kadanoff-Baym (KB)
form (\ref{eq:general_transport}).

We recall that the consistent Botermans-Malfliet form
(\ref{eq:general_transport_bm}) has been used as a starting point
for the  off-shell description of realistic heavy-ion collisions
\cite{caju1,caju2,caju3}. Together with an extended  test-particle
ansatz for $i G^<$ this equation allows for a convenient solution
of the generalized transport equation in phase-space including the
dynamics of the spectral
function. However, it is presently not clear in how far the
generalized transport equations in KB-form
(\ref{eq:general_transport}) or in BM-form
(\ref{eq:general_transport_bm}) reproduce the same dynamics as for
the full Kadanoff-Baym theory (\ref{kabaeqcs}). Moreover, the differences in the
time evolution of nonequilibrium systems between the KB-form
(\ref{eq:general_transport}) and BM-form
(\ref{eq:general_transport_bm}) are not known either. We will thus
perform an explicit comparison between the different limits for
$\phi^4$-theory in 2+1 dimensions in the following Section. For a
related and more detailed comparison between the solution of the full KB equations
with those from the on-shell Boltzmann limit we refer the reader
to Section V of Ref. \cite{Juchem03}.

\section{\label{sec:gradient_dynamics} Dynamics within the
Generalized Transport Equation}

In this Section we will perform numerical studies of the dynamics
inherent in the generalized transport equations derived in Section
2.2 in comparison to solutions of the full Kadanoff-Baym equations
(\ref{kabaeqcs}) for the $\phi^{4}$-theory in 2+1 space-time
dimensions within the three-loop approximation for the effective
action carried out in Ref. \cite{Juchem03}. This fixes the
self-energies $\bar{\Sigma}^{\delta}$ and
$i\bar{\Sigma}^{\gtrless}$ in (\ref{eq:general_transport}) and
(\ref{eq:general_mass}) to be the same as in the case of the full
Kadanoff-Baym theory given by the tadpole (\ref{tadpole_cs}) and
the sunset (\ref{sunset_cs}) contributions, respectively. We
recall that both self-energies are ultraviolet divergent and have
to be renormalized by proper counter terms as shown in Appendix B
of Ref. \cite{Juchem03}. In the following we will use the
renormalized mass $m$=1, which implies that times are given in
units of the inverse mass. Accordingly, the coupling $\lambda$ is
given in units of mass $m$ such that $tm$ and $\lambda/m$ are
dimensionless.

For the first investigation we
concentrate on the dynamics of the generalized transport equation
(\ref{eq:general_transport}). As in
Ref. \cite{Juchem03} we restrict ourselves to homogeneous
systems in space. Consequently the derivatives with respect to the
mean spatial coordinate $\vec{x}$ vanish, such that the
generalized transport equation (\ref{eq:general_transport})
reduces to \\
\bea
\label{eq:homo_general_transport}
2\,p_0\:\partial_{t} \: i\bar{G}^{<}
\, - \,
\{ \, \bar{\Sigma}^{\delta} \!+\! Re\,\bar{\Sigma}^{R} \, ,
   \, i \bar{G}^{<} \, \}
\, - \,
\{ \, i\bar{\Sigma}^{<} \, , \, Re\,\bar{G}^{R} \, \}
\: = \:
i\bar{\Sigma}^{<} \; i\bar{G}^{>}
\, - \,
i\bar{\Sigma}^{>} \; i\bar{G}^{<} \;\;\;\;\;\;
\eea\\
with the simplified Poisson brackets
(for arbitrary functions
$\bar{F}, \bar{G}$) \\
\bea
\label{eq:def_homo_poisson}
&&
\{ \, \bar{F}(\vec{p},p_0,t) \, , \, \bar{G}(\vec{p},p_0,t) \, \}
\\[0.4cm]
&& \qquad \qquad \; = \;
\partial_{p_0} \,\bar{F}(\vec{p},p_0,t) \;
\partial_{t}   \,\bar{G}(\vec{p},p_0,t) \; - \;
\partial_{t}   \,\bar{F}(\vec{p},p_0,t) \;
\partial_{p_0} \,\bar{G}(\vec{p},p_0,t) \; . \phantom{aaa}
\nn
\eea\\
In order to obtain a numerical solution the Kadanoff-Baym equation
(\ref{kabaeqcs}) as well as the generalized transport equations
(\ref{eq:general_transport}) are transformed to momentum space.
For the numerical solution  we have developed a flexible and
accurate algorithm, which is described in more detail in Appendix
A of Ref. \cite{Juchem03}. Furthermore,  since in 2+1 space-time
dimensions both self-energies (\ref{tadpole_cs}),(\ref{sunset_cs})
are ultraviolet divergent, they have to be renormalized by
introducing proper counter terms. The details of the
renormalization scheme are given in Appendix B of Ref.
\cite{Juchem03} as well as a numerical proof for the convergence
in the ultraviolet regime.  For the actual numerical integration
of the generalized transport equation we use the same grid in
momentum space as employed for the full KB-theory in Ref.
\cite{Juchem03}.

\subsection{\label{sec:gradient_initial} The Initial State}

In order to specify the problem for the first order
transport equation in time, the initial state has to be fixed,
i.e. the initial Green function
$i\bar{G}^{<}(\vec{p},p_0,t=0)$ has to be specified for all momenta
$\vec{p}$ and all energies $p_0$.
To this aim, we first assume a given initial distribution function
$n(\vec{p},t\!=\!0)$ in momentum space equivalent to those used in
the investigation of the full Kadanoff-Baym theory in Ref. \cite{Juchem03}
in order to allow for a comparison of the two schemes.
Next, the complete initial phase-space distribution function
$\bar{N}(\vec{p},p_0,t\!=\!0)$ has to be specified as a function of
the energy $p_0$ as well.
We achieve this by assuming $\bar{N}$ to be
constant in the energy variable $p_0$ as a starting point.
In order to determine the initial Green function
we employ the self-consistent iteration procedure
for the full spectral function as described in Appendix D of Ref. \cite{Juchem03}
 using the specified nonequilibrium
distribution function $\bar{N}$.
The iteration process then yields the fully self-consistent spectral
function $\bar{A}(\vec{p},p_0,t\!=\!0)$ for this initial distribution and
thus determines the initial Green function via \\
\bea
i\bar{G}^{<}(\vec{p},p_0,t\!=\!0) \; = \;
\bar{N}(\vec{p},p_0,t\!=\!0) \; \bar{A}(\vec{p},p_0,t\!=\!0) \: .
\eea\\
In principle one might choose the initialization --
in particular the energy dependence of the distribution functions --
in various ways.
We adopt here the simple description introduced above since it
approximately reproduces the same equal-time (i.e. energy integrated)
initial Green function as in the full Kadanoff-Baym calculation in Ref. \cite{Juchem03}.
However, with increasing coupling strength $\lambda$ the spectral
function achieves a considerable width and the on-shell energies of the
momentum modes are slightly shifted.
In this case the equal-time Green function deviates
somewhat in the low momentum region from the `quasiparticle'-like
initialization used for the Kadanoff-Baym calculation (cf. Ref. \cite{Juchem03}).

\begin{figure}[htp]
\centerline{\psfig{file=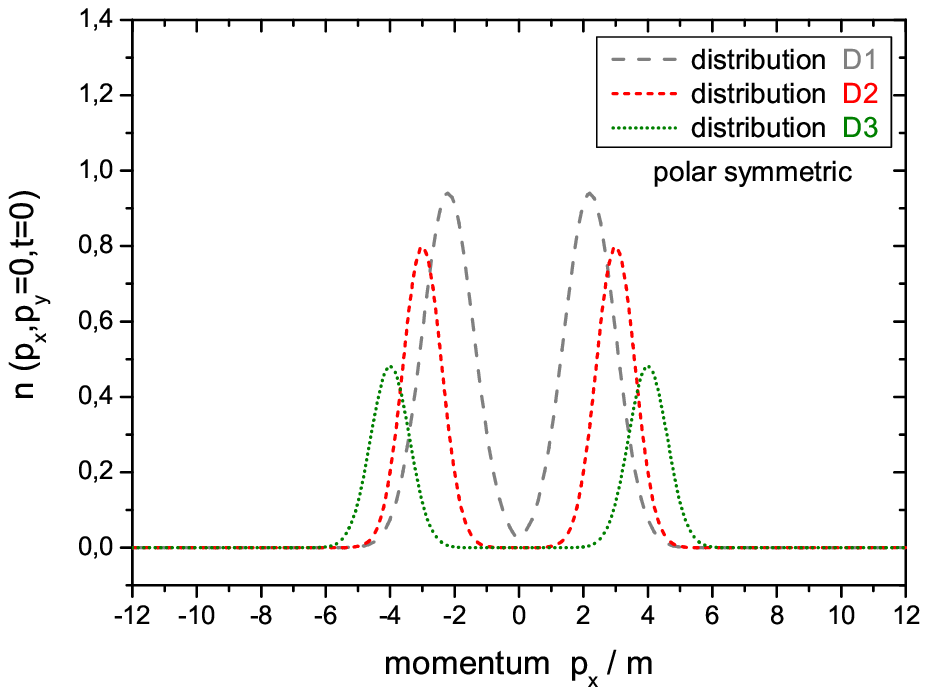,width=12cm}} \vspace{-1.0cm}
\caption{\label{fig:4phd_gradient_inin} Initial momentum
distributions for the initializations D1, D2 and D3. For the polar
symmetrical settings in momentum space the distributions are shown
as function of $p_x$ (for $p_y=0$).}
\centerline{\psfig{file=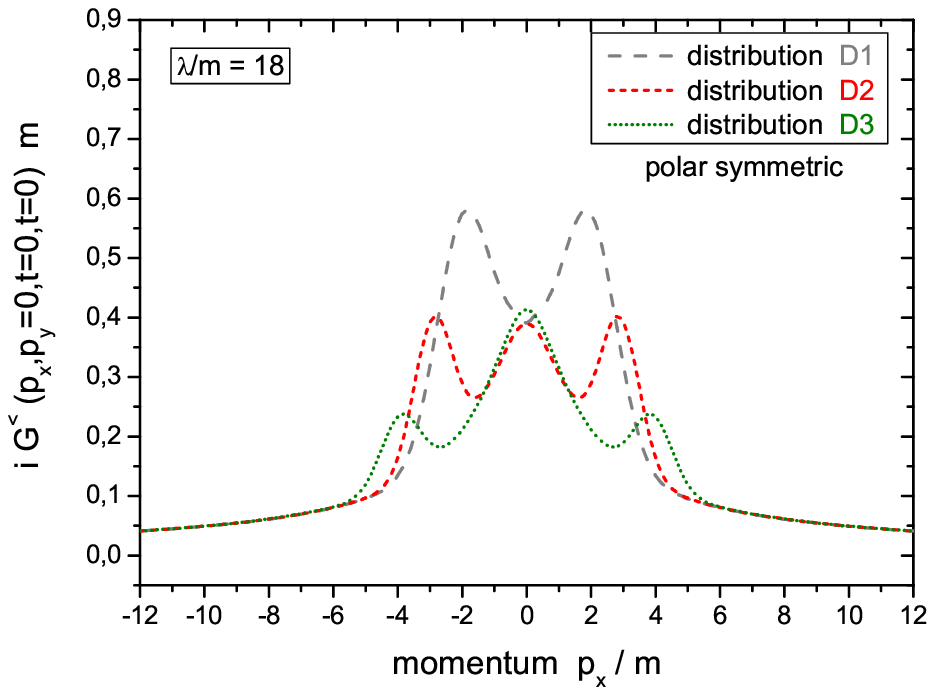,width=12cm}} \vspace{-1.0cm}
\caption{\label{fig:4phd_gradient_inig} Initial equal-time Green
functions for the initializations D1, D2 and D3. The equal-time
Green functions are displayed for polar symmetric systems as a
function of $p_x$ (for $p_y=0$) with coupling constant $\lambda/m
= 18$.}
\end{figure}

The initial distribution functions for the following studies
are shown in Fig. \ref{fig:4phd_gradient_inin} for
polar symmetric configurations as a function of the momentum coordinate
$p_x$ for $p_y$ = 0.
They are identical to those used in Ref. \cite{Juchem03} within the full Kadanoff-Baym
theory.
The resulting initial equal-time Green functions -- for the
coupling constant $\lambda/m = 18$ -- are displayed in
Fig. \ref{fig:4phd_gradient_inig}.
By comparison with the initial Green functions of the
Kadanoff-Baym calculation (c.f. lower part of Fig. 3 in Ref. \cite{Juchem03})
we find tiny deviations in the region of small absolute momenta.
This is, of course, a consequence of the large coupling constant
$\lambda$
employed, which we  use in order to compare to the calculations
within the full quantum evolution for the same coupling strength.
Since the initial states are very close to those
used for the Kadanoff-Baym theory, we will also denote them
as initializations D1, D2 and D3 as in Ref. \cite{Juchem03}.

The advantage of the initialization prescription introduced above
is that the actual spectral function -- directly obtained by
$\bar{A} = i\bar{G}^{>} - i\bar{G}^{<}$ from the Green functions
-- complies with the one determined from the self-energies
(\ref{eq:specorder0}) in accordance with the first order gradient
expansion scheme. During the nonequilibrium time evolution this
correspondence is maintained since the {\em analytic} expression
for the spectral function already is a solution of the generalized
transport equation itself. Furthermore, the real part of the
retarded Green function, that enters the peculiar second Poisson
bracket on the l.h.s. in (\ref{eq:general_transport}), can be
taken in the first order scheme (\ref{eq:regretorder0}) which
simplifies the calculations considerably.

We mention that other prescriptions are also possible for the calculation
of $Re\,\bar{G}^{R}$:\\
1) One can determine it directly from
$i\bar{G}^{\gtrless}$ by a Fourier technique similar to one used
in the calculation of the self-consistent spectral functions
in Appendix D of Ref. \cite{Juchem03}.
Here the real part of the retarded Green function is obtained
via inverse Wigner transformation with respect
to the energy, multiplication by the  $\Theta$-function in relative time and
transformation back to phase-space. \\
2) One can also use dispersion relations with the spectral
function to specify $Re\,\bar{G}^{R}$. \\
As we have checked in our actual simulations all prescriptions lead to
practically identical results.

\subsection{\label{sec:gradient_equilibration} Numerical Study of
Equilibration}

\begin{figure}[th]
\centerline{\psfig{file=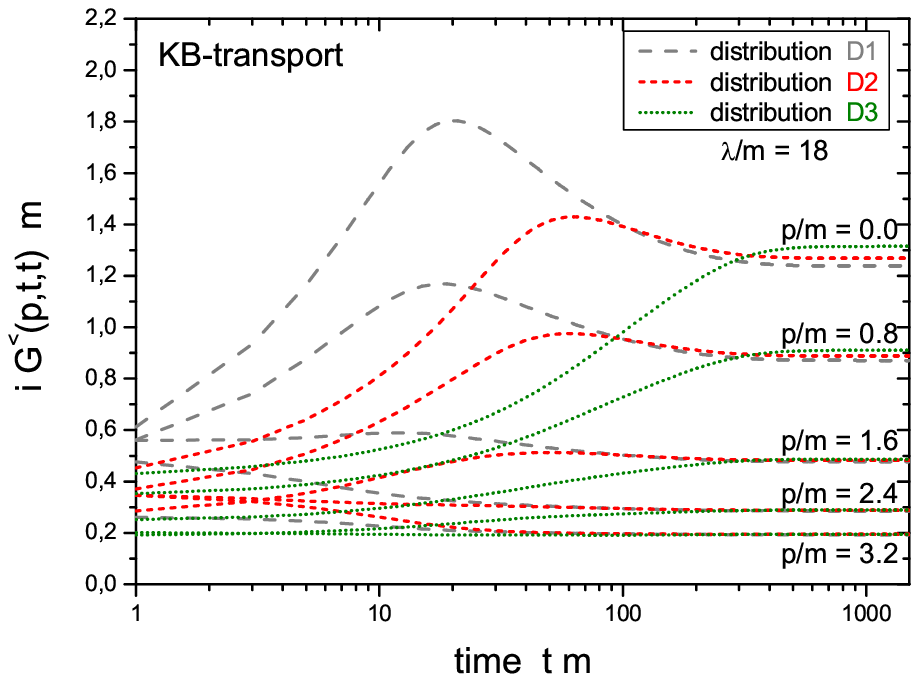,width=12cm}} \vspace{-1.0cm}
\centerline{\psfig{file=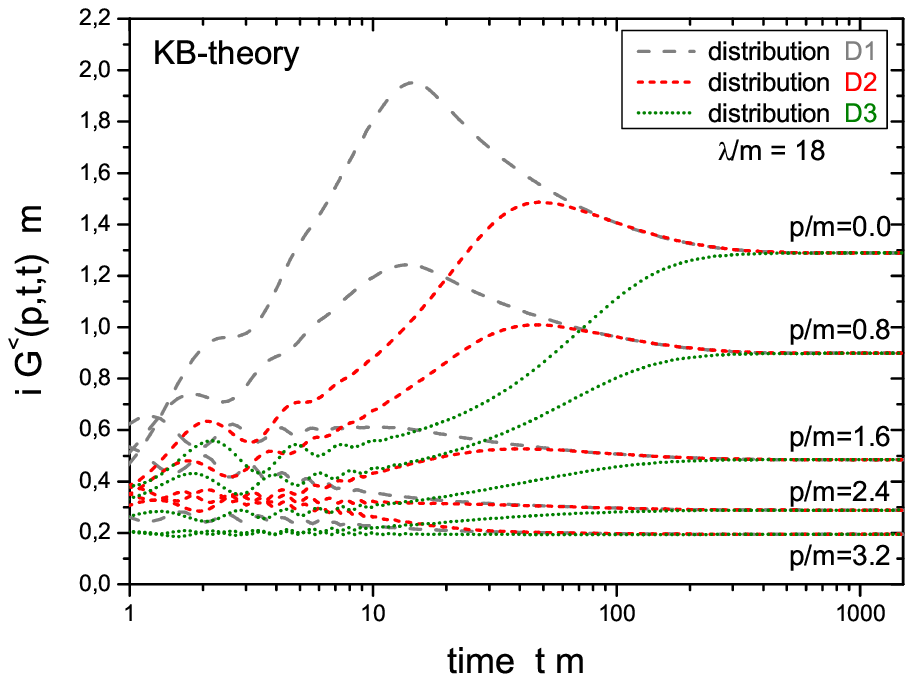,width=12cm}}
\caption{\label{fig:4phd_gradient_modes} Upper part: Evolution of
several momentum modes $|\,\vec{p}\,|/m =$ 0.0, 0.8, 1.6, 2.4,
3.2, 4.0 of the equal-time Green function on a logarithmic time
scale for the different initializations D1, D2 and D3 for the
generalized transport equation
(\protect\ref{eq:general_transport}). Lower part: same as above
but for the full Kadanoff-Baym equation (\protect\ref{kabaeqcs}).}

\end{figure}

Now we turn to the actual solutions of the
generalized transport equation in the KB form (\ref{eq:general_transport}).
In Fig. \ref{fig:4phd_gradient_modes} (upper part) we show the time
evolution of the equal-time Green function
$iG^{<}(|\,\vec{p}\,|,t,t)$ for the
polar symmetric initial states D1, D2 and D3 as specified in Fig. 2.
We have displayed several momentum modes
$|\,\vec{p}\,|/m =$ 0.0, 0.8, 1.6, 2.4, 3.2, 4.0 of the equal-time
Green function on a logarithmic time scale.
As in the full Kadanoff-Baym theory (lower part) we find that for all initializations
the quantum system approaches a stationary state for
$t \rightarrow \infty$, i.e. all momentum modes approach a constant.
However, the respective momentum modes of the different
initializations do not achieve identical values for
$t \rightarrow \infty$, as seen
in particular for the low momenta $|\,\vec{p}\,|/m =$ 0.0, 0.8 in
 Fig. \ref{fig:4phd_gradient_modes} (upper part).
This is not surprising since the various initializations --
obtained within the self-consistent scheme described above -- do
not correspond to exactly the same energy. This is why the
respective long-time limits differ slightly. The small difference
in energy is, of course, most prominently seen in the low momentum
(energy) modes. Moreover, the dynamics within the generalized
transport equation (\ref{eq:general_transport}) is in general very
similar to the full Kadanoff-Baym theory (lower part). For all
three initial states we find (apart from the very initial phase $t
\cdot m <$ 5) the same structures during the equilibration
process. In particular for the initializations D1 and D2 the
characteristic overshooting for the low momentum modes is seen as
in the full quantum evolution, which does not show up in solutions
of the corresponding on-shell Boltzmann limit. Since in the
Boltzmann limit a strictly monotonous evolution of the momentum
modes is seen  (cf. Ref. \cite{Juchem03}) this overshooting has to
be attributed to an off-shell quantum effect. Even the positions
of the maxima are in a comparable range: For the initialization D1
they are shifted to slightly larger times and are a little bit
lower than in the full calculation; the same holds for the initial
state D2. The initial distribution D3 yields a monotonous
behaviour for all momentum modes within the generalized transport
formulation which is again in a good agreement with the full
dynamics.

Some comments are worthwhile with respect to the comparison
performed above: The spectral function in the Kadanoff-Baym
calculation is completely undetermined in the initial state; it
develops during the very early phase to an approximate form (which
in the following still evolves in time). In contrast to this, the
spectral function in the generalized transport formulation
(\ref{eq:general_transport}) has a well-defined structure already
from the beginning. This principle difference results from the
fact, that in the Kadanoff-Baym case we deal with a true initial
value problem in the two time directions ($t_1,t_2$). Moreover,
the spectral distribution for very early times depends on the
specific initial conditions adopted (cf. Appendix C of Ref.
\cite{Juchem03}). Additionally, the relative time integral in
($t_1-t_2$) -- to obtain the spectral function in energy $p_0$ by
Wigner transformation -- is very small in the initial phase as
discussed in Ref. \cite{Juchem03}.  Consequently, the spectral
shape in Wigner space is determined  by the finite integration
interval in time rather than by the interactions itself. On the
other hand, we have used an infinite relative time range in
deriving the generalized transport equation within the first order
gradient expansion. Thus in this case we deal with a completely
resolved spectral function already at the initial time, that
clearly exhibits the physics incorporated. This demonstrates why
both approaches can only be compared to a certain extent for the
very early times.

Nevertheless, the inclusion of the dynamical spectral function in
terms of the generalized transport equation surpasses the shortcomings
of the quasiparticle Boltzmann limit discussed in Section V
and Appendix E of Ref. \cite{Juchem03}.
Whereas the latter approach leads to a strictly monotonous evolution of the
momentum modes, the inclusion of quantum effects in terms of a semiclassical
approximation correctly yields the overshooting effects as in
the solution of the full Kadanoff-Baym equation.

Finally, concentrating on the very early time behaviour, we find a
significant difference between the full and the approximate
dynamics in the gradient scheme (\ref{eq:general_transport}). For
the generalized transport equation we find a monotonous evolution
of the equal-time Green function momentum modes, whereas strong
oscillations are observed in the initial phase for the solution of
the full Kadanoff-Baym theory (see also Fig. 4 of Ref.
\cite{Juchem03}). Thus, with respect to the early time behaviour
the generalized transport equation behaves much more like the
Boltzmann approximation, which is a first order differential
equation in time as well. However, the Kadanoff-Baym evolution is
given by an integro-differential equation of second order in time.
In this case the phase correlations between the Green functions
$G^{<}_{\phi \phi},\,G^{<}_{\pi \phi},\,G^{<}_{\pi \pi}$ are kept
and the instantaneous switching-on of the interaction results in
an oscillatory behaviour of the single momentum modes. We mention
that correlations build up very rapidly in the very early phase of
the evolution in the full KB theory (cf. Fig. 6 of Ref.
\cite{Juchem03}) that may be attributed to this oscillatory
behaviour. When including the collisional self-energies, i.e. on
the three-loop level for the effective action, these oscillations
are damped in time typically with the respective on-shell width
for given momentum. If only the tadpole term is included, these
oscillations are not damped and maintain forever (see also Fig. 4
of Ref. \cite{Juchem03}). Therefore, the origin of the oscillatory
behaviour can be traced back to the order of the underlying
differential equation and the rapid build-up of correlations that
have not been incorporated in the initial conditions for the full
KB-theory.

\subsection{\label{sec:gradient_spectralevolution} Evolution of the
Spectral Function}

\begin{figure}[t]
\centerline{\psfig{file=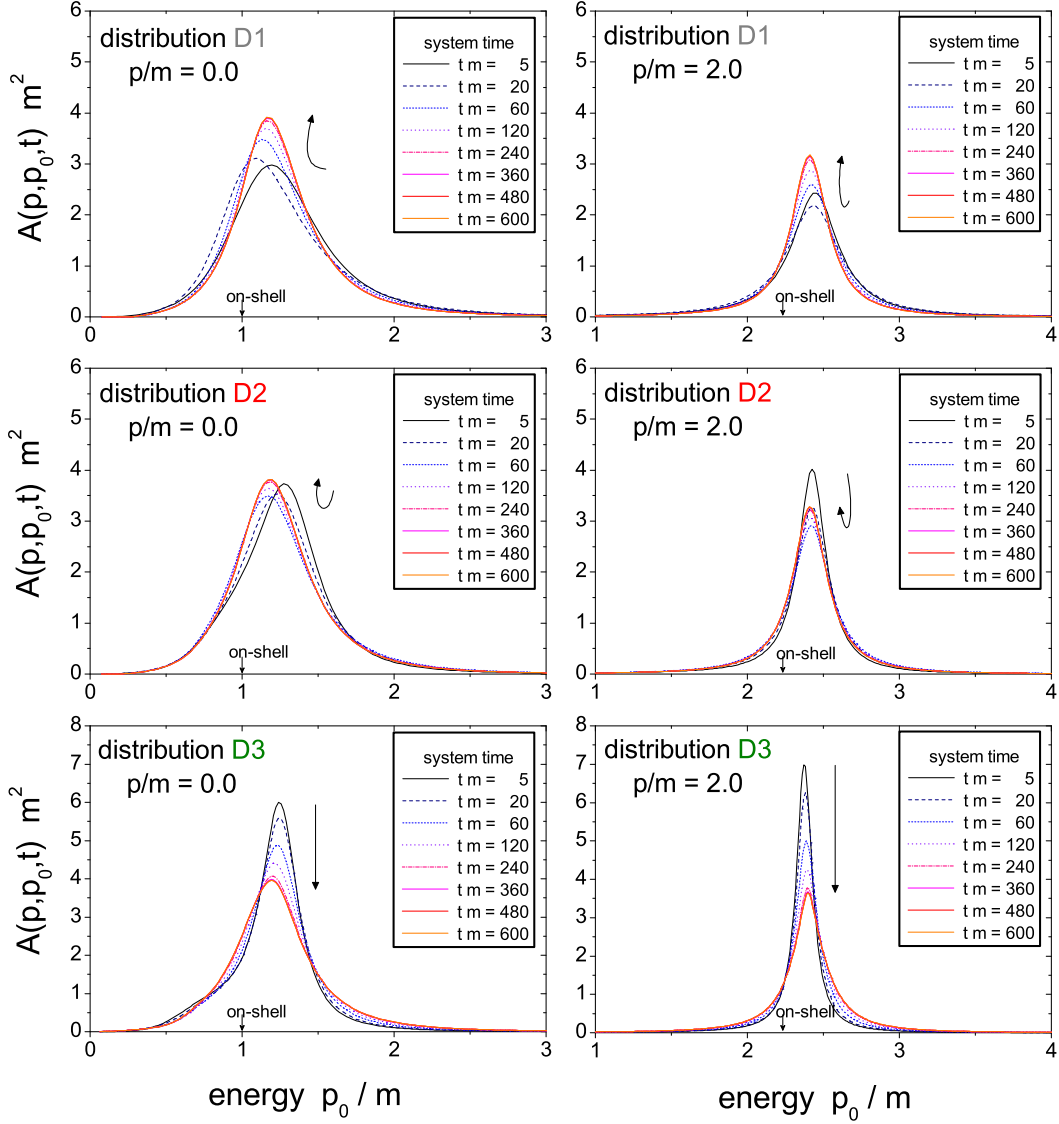,width=16cm}} \vspace{-0.1cm}
\caption{\label{fig:4phd_gradient_spec} Time evolution of the
spectral function $\bar{A}(\vec{p},p_0,t)$ for the initial
distributions D1, D2 and D3 (from top to bottom) with coupling
constant $\lambda / m =$ 18 for the two momentum modes
$|\,\vec{p}\,| / m =$ 0.0 (l.h.s.) and $|\,\vec{p}\,| / m =$ 0.0
(r.h.s.). The spectral function from the transport eq.
(\ref{eq:general_transport}) is shown for times $t \cdot m =$ 5,
20, 60, 120, 240, 360, 480, 600 as indicated by the different line
types.} \vspace*{-0.7cm}
\end{figure}

Since the Green functions develop in time also
the spectral properties of the system change as well.
In Fig. \ref{fig:4phd_gradient_spec} the
time evolution of the spectral functions for the
initializations D1, D2 and D3 within the gradient scheme
are displayed.
We focus on the spectral functions for two particular
momentum modes
$|\,\vec{p}\,| / m =$ 0.0 (l.h.s.) and
$|\,\vec{p}\,| / m =$ 2.0 (r.h.s.) for various system times
$t \cdot m =$ 5, 20, 60, 120, 240, 360, 480, 600
up to the long-time limit.
This representation corresponds to Fig. 7 in Ref. \cite{Juchem03},
where the respective evolution of the spectral function is studied
for the full Kadanoff-Baym theory.
We find that the time evolution of the spectral functions
obtained from the generalized transport equation
(\ref{eq:general_transport}) is very similar to the one from
the full quantum calculation (see below).
The zero-mode spectral function for the initial distribution
D1 becomes sharper with time and is moving to slightly
higher energies.
The opposite characteristics is observed for the zero-mode
spectral function for the initialization D3, which broadens with
time (reducing the peak correspondingly)
and slowly shifts to smaller energies.
Together with the weak evolution for the distribution D2
(which only slightly broadens at intermediate times and
returns to a narrower shape at smaller energies in the long-time limit)
the evolution of all three initializations in the semiclassical
approximation is well comparable to the full Kadanoff-Baym dynamics
(cf. Fig. 7 in Ref. \cite{Juchem03}).
Furthermore, the maxima of the zero-mode spectral functions
are located above the bare mass (as indicated by the
on-shell arrow) for all initial states during the time
evolution.

The spectral functions for the momentum mode
$|\,\vec{p}\,| / m =$ 2.0 are in a good agreement with
the Kadanoff-Baym dynamics as well.
Again we observe -- for the initial distribution D1 --
a narrowing of the spectral function,
while for D3 the spectral function broadens with time.
Moreover, the width of the spectral function starting
from distribution D2 shows a non-monotonous behaviour
with a maximum at intermediate times.

\begin{figure}[t]
{\psfig{file=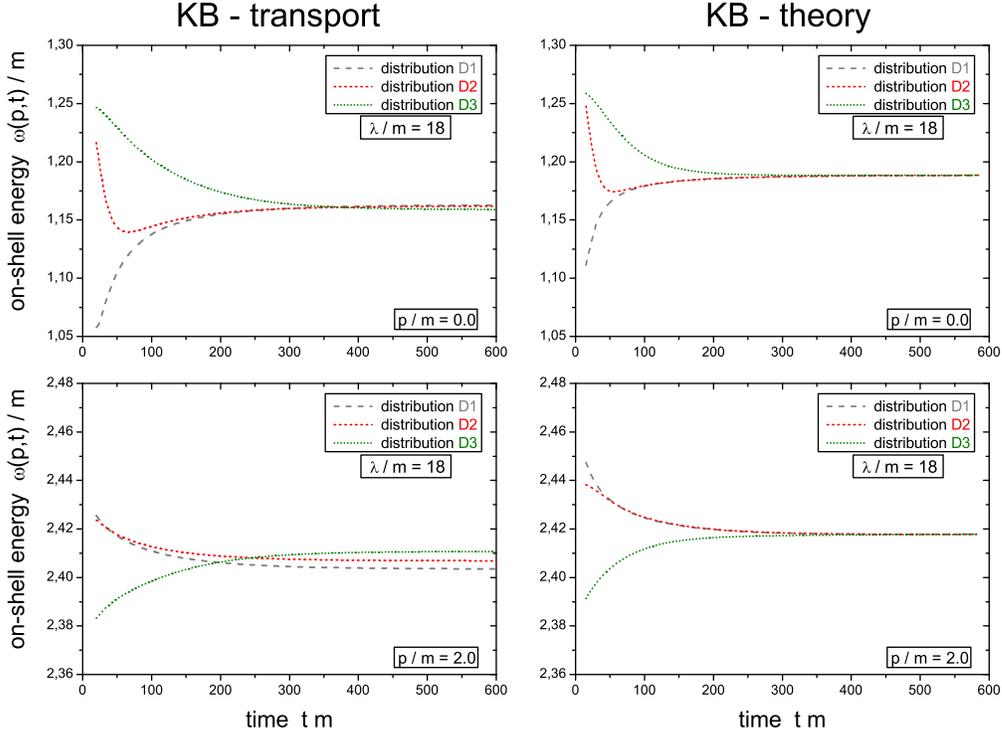,width=15cm}} \vspace{-0.1cm}
\caption{\label{fig:4phd_gradient_ose01} Time evolution of the
on-shell energies $\omega(\vec{p},t)$ of the momentum modes
$|\,\vec{p}\,|/m = 0.0$ and $|\,\vec{p}\,|/m = 2.0$ for the
different initializations D1, D2 and D3 with $\lambda / m = 18$ in
the semiclassical KB limit (\ref{eq:general_transport}) (l.h.s.).
The on-shell self-energies are extracted from the maxima of the
time-dependent spectral functions. The respective results
 from the full Kadanof-Baym theory are displayed on the r.h.s.}
\vspace{-0.5cm}
\end{figure}

In order to study the dynamics of the spectral function
in a more quantitative manner we display in
Fig. \ref{fig:4phd_gradient_ose01}
the time evolution of the on-shell energies (as derived from the
maxima of the spectral function) for the
momentum modes $|\,\vec{p}\,| / m =$ 0.0 (upper plot)
and 2.0 (lower plot) for the initializations D1, D2 and D3
with $\lambda / m =$ 18 (l.h.s.).
By comparison with the corresponding results from the Kadanoff-Baym
theory (r.h.s.) we observe a close similarity of the
evolutions within the full and the semiclassical KB scheme.
The effective mass of the zero momentum
mode decreases for initialization D3, passes a minimum for D2
and increases for the initial state D1.

As familiar from the Kadanoff-Baym calculations in
Section IV.B of \cite{Juchem03} the behaviour of
the on-shell energies is different for higher momentum modes.
We find for the momentum mode $|\,\vec{p}\,| / m =$ 2.0 a
monotonous decrease of the on-shell energy for the
initializations D1 and D2 and an increase for distribution D3.
Altogether, the evolution of the on-shell energies for the higher
modes is rather moderate compared to the lower ones in
accordance with the dominant momentum contribution and the weakening of
the retarded self-energy for higher energy  modes.

Finally, the on-shell energies approach a stationary state
for all modes and all initializations.
However, the long-time limit of the equal momentum modes is
not exactly the same for all initial distributions D1, D2 and D3.
As discussed above this small difference can be traced back to the specific initial
state generation from the given momentum distribution.

\begin{figure}[t]
{\psfig{file=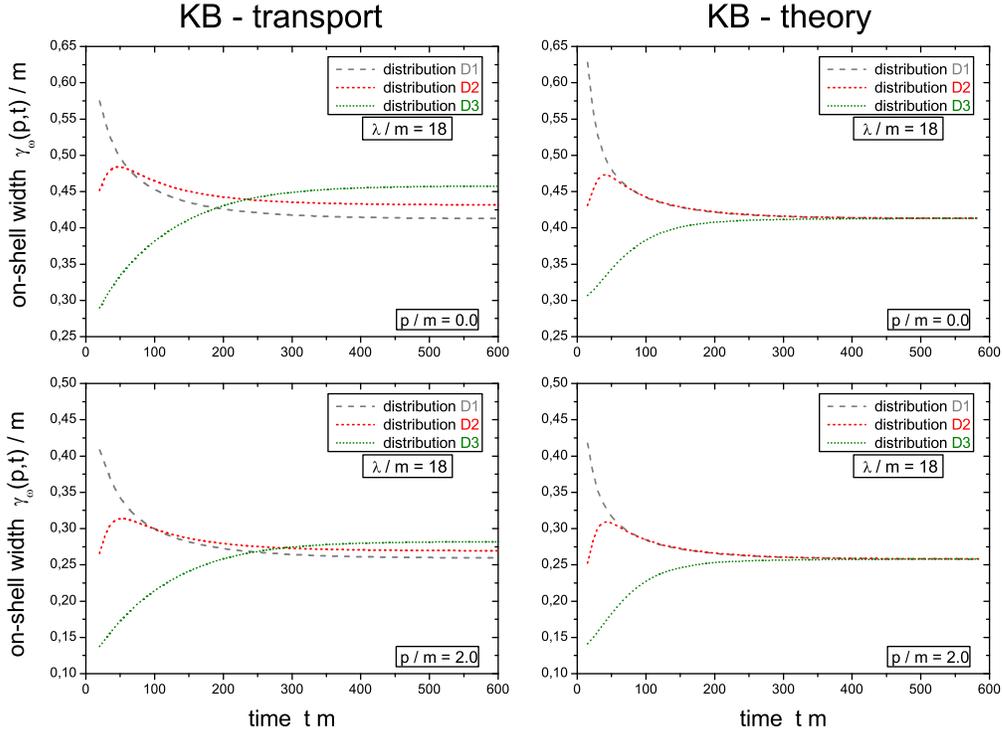,width=15cm}} \vspace{-0.1cm}
\caption{\label{fig:4phd_gradient_osw01} Time evolution of the
on-shell widths $-
Im\,\bar{\Sigma}^{R}(\vec{p},\omega(\vec{p},t),t) /
\omega(\vec{p},t)$ of the momentum modes $|\,\vec{p}\,|/m = 0.0$
and $|\,\vec{p}\,|/m = 2.0$ for the different initializations D1,
D2 and D3 with $\lambda / m = 18$ in the semiclassical transport
eq. (\ref{eq:general_transport}) (l.h.s.). The respective results
 from the full Kadanof-Baym theory are displayed on the r.h.s.}
\vspace{-0.5cm}
\end{figure}

Next we consider the time evolution of the on-shell width
as determined by the imaginary part of the retarded self-energy
at the maximum position of the spectral function.
In Fig. \ref{fig:4phd_gradient_osw01} (l.h.s.) the on-shell width
is displayed for the two momentum modes
$|\,\vec{p}\,| / m =$ 0.0 and
$|\,\vec{p}\,| / m =$ 2.0 for all three initial distributions
D1, D2 and D3 with $\lambda / m = $18 as a function of time.
For both momentum modes the on-shell width increases for the
distribution D3, while it has a maximum at intermediate times
($t \cdot m \approx$ 40) for the initialization D2.
Thus the results -- together with the reduction of the on-shell
width for both momentum modes for the initialization D1 -- is in
good agreement with the results obtained for the full Kadanoff-Baym
theory (r.h.s.).
However, the stationary values for the on-shell widths
deviate again slightly in accordance with the different preparation of the
initial state in the gradient scheme.

In summarizing  we find that the main characteristics
of the full quantum evolution of the spectral function are
maintained in the semiclassical transport equation
(\ref{eq:general_transport}) as well.
This includes the evolution of the on-shell energies as well
as the width of the spectral function.
Since the generalized transport equation is formulated directly
in Wigner space one has access to the spectral properties
at all times, whereas the very early times in
the Kadanoff-Baym case have to be excluded
due to the very limited support in the relative time
interval ($t_1 - t_2$) for the Wigner transformation.

\subsection{\label{sec:gradient_stationary} Stationary State of the
Semiclassical Evolution}

\begin{figure}[htp]
{\psfig{file=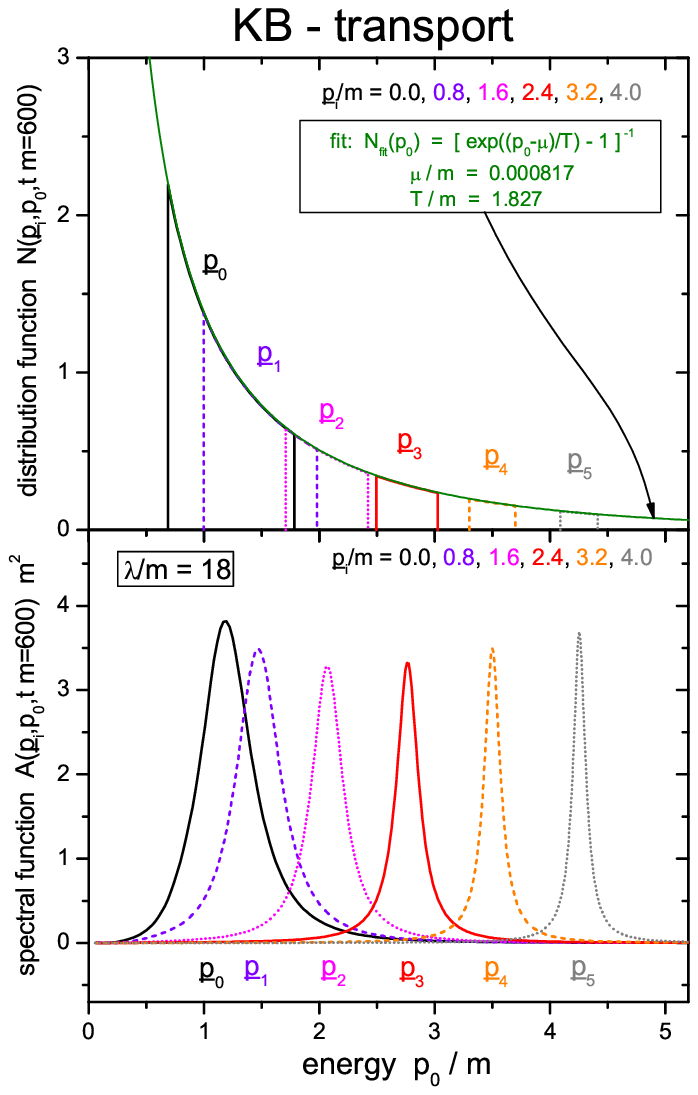,width=7.5cm}} \hspace{-0.2cm}
{\psfig{file=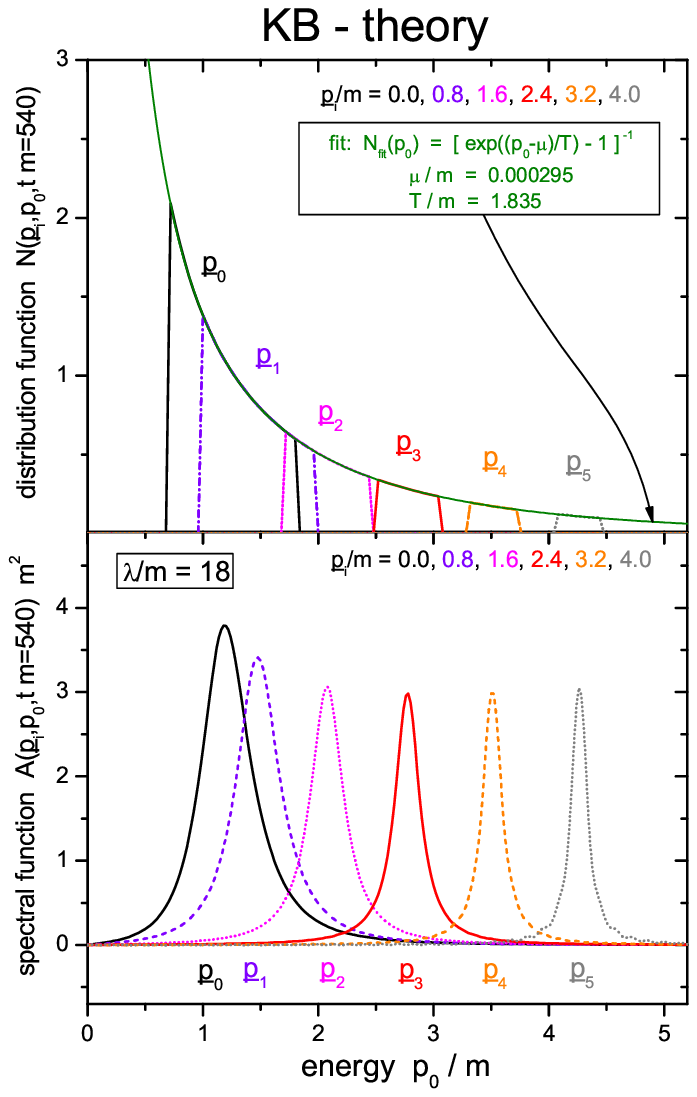,width=7.5cm}} \vspace{1.0cm}
\caption{\label{fig:4phd_gradient_kmsna} Spectral function
$\bar{A}$ for various momentum modes as a function of energy $p_0
/ m$ at the late time $t \cdot m =$ 600 (lower part) for initial
distribution D2 with coupling constant $\lambda/m = 18$ in the
semiclassical transport limit (\ref{eq:general_transport})
(l.h.s.). The corresponding distribution function $\bar{N}$ (at
the same time for the same momentum modes) is shown in the upper
part. All momentum modes can be fitted for all energies by a
single Bose function of temperature $T / m =$ 1.827 and a chemical
potential $\mu / m$ close to zero. The respective results
 from the full Kadanof-Baym theory are displayed on the r.h.s.}
\end{figure}

As we have observed in the previous Subsections the evolution
within the generalized transport equation
(\ref{eq:general_transport}) leads to a stationary state
for all three different initializations D1, D2 and D3.
Thus we turn to the investigation of this long-time limit itself, here
in particular for the initialization D2.
In Fig. \ref{fig:4phd_gradient_kmsna} (l.h.s.) we show the distribution
function $\bar{N}$ of various momentum modes
$|\,\vec{p}\,| / m =$ 0.0, 0.8, 1.6, 2.4, 3.2, 4.0 for large times
($t \cdot m$ = 600) as derived from the Green function itself and the
spectral function via the relation $\bar{N} = i\bar{G}^{<} / \bar{A}$.
The distribution function for a given momentum mode is calculated
for all energies $p_0$ where the corresponding spectral function
-- as displayed in the lower part of Fig.
\ref{fig:4phd_gradient_kmsna} -- exceeds a value of 0.5. Since the
width of the late time spectral function decreases with increasing
momentum, the energy range for which the distribution function is
shown, is smaller for larger momentum modes. We find, that all
momentum modes of $\bar{N}$ can be fitted at all energies by a
single Bose function with a temperature $T / m =$ 1.827 and a very
small chemical potential $\mu / m =$ 0.000817. Thus the
generalized transport formulation (\ref{eq:general_transport})
leads to a complete (off-shell) equilibration of the system very
similar to the solution of the full Kadanoff-Baym equation (r.h.s.
of Fig. \ref{fig:4phd_gradient_kmsna}). Furthermore, the long-time
limit of the semiclassical time evolution exhibits a vanishing
chemical potential $\mu / m$ in accordance with the properties of
the neutral $\phi^4$-theory. This might have been expected since
in the generalized transport equation particle number
non-conserving processes of the type $1 \leftrightarrow 3$ --
which lead to the decrease of the chemical potential -- are
included by means of the dynamical spectral function. Thus the
semiclassical approximation (\ref{eq:general_transport}) solves
the problems within the Boltzmann limit, which does not yield a
 relaxation of the chemical potential, since only on-shell
$2 \leftrightarrow 2$ transitions of quasiparticles are taken into
account as demonstrated in Section V of Ref. \cite{Juchem03}.

\begin{figure}[th]
\centerline{\psfig{file=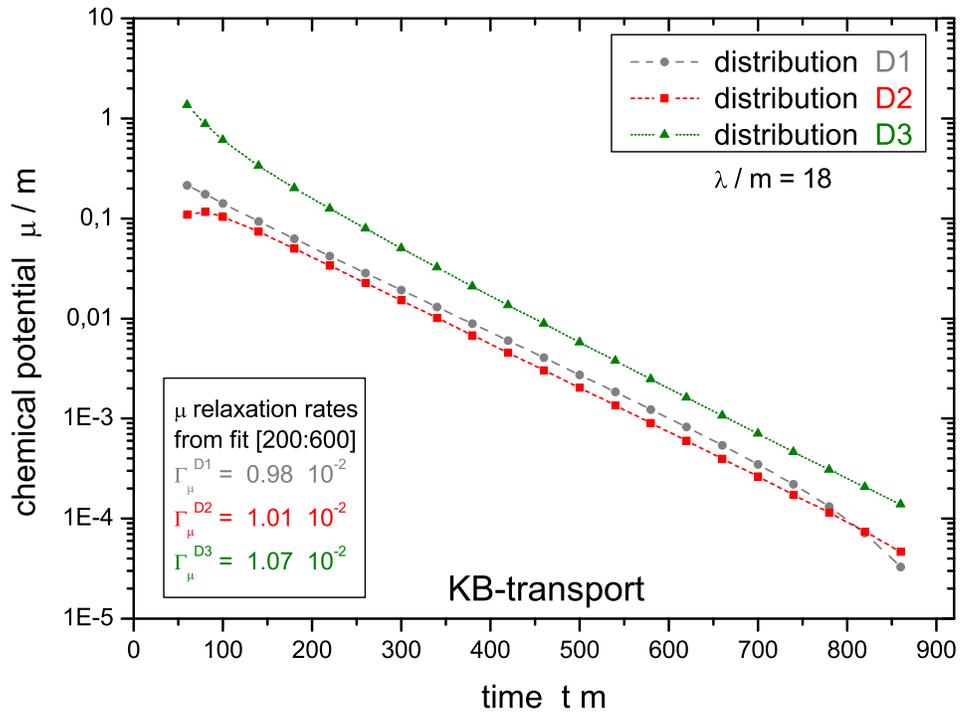,width=15cm}} \vspace{1.0cm}
\caption{\label{fig:4phd_gradient_mul18} Relaxation of the
chemical potential $\mu / m$ for the three initial distributions
D1, D2, D3 within the generalized transport equation
(\ref{eq:general_transport}) in a logarithmic repesentation as a
function of time. The decrease is approximately exponential in
time $\propto \exp(-\Gamma_{\mu} \cdot t)$ and the relaxation rate
$\Gamma_{\mu}$ is practically equal for all initializations.}
\end{figure}

After observing, that the chemical potential decreases to zero
in the long-time limit, it is interesting to study the relaxation
process itself.
The relaxation of the chemical potential $\mu / m$ is shown for the
three different initializations D1, D2 and D3 with coupling constant
$\lambda / m =$ 18 in Fig. \ref{fig:4phd_gradient_mul18}.
We see -- as in the case of the Kadanoff-Baym evolution -- that
all initial states show an approximately exponential decrease in time.
The relaxation rates -- as determined from the slope of the exponential
decline -- are also approximately the same for all distributions.
They are given by
$\Gamma^{D1}_{\mu} \approx 0.98 \cdot 10^{-2}$ for distribution D1,
$\Gamma^{D2}_{\mu} \approx 1.01 \cdot 10^{-2}$ for distribution D2 and
$\Gamma^{D3}_{\mu} \approx 1.07 \cdot 10^{-2}$ for distribution D3.
Thus the relaxation rates are in the same range as those
found within the full Kadanoff-Baym theory (Section IV.D in Ref. \cite{Juchem03}).
This is exactly the result one expects from the analytical estimate
for the chemical potential relaxation rate.
In Section IV.D of Ref. \cite{Juchem03} we have found that the relaxation rate
$\Gamma_{\mu}$ can be explained within a linearized evolution equation
including only equilibrium properties, i.e. the equilibrium spectral and
(Bose) distribution function.
It is appropriate for small deviations from the equilibrium state
in terms of the chemical potential $\mu$.
In the present case of the generalized transport equation
(\ref{eq:general_transport}) we encounter exactly the same situation.
We know from the validity of the estimate that a linearized
description is meaningful.
Thus the evolution within the first order gradient equation
should yield a comparable result as long as the equilibrium
properties are about equal.
This is indeed the case since the final temperatures for the various
initial states are approximately the same
(also compared to the Kadanoff-Baym case).
They are given by
$T^{D1}_{eq} / m \approx 1.819$,
$T^{D2}_{eq} / m \approx 1.827$,
$T^{D3}_{eq} / m \approx 1.834$ compared to $T_{eq} / m \approx 1.835$
in the full Kadanoff-Baym case.
Consequently, the same similarity holds for the spectral
function, which is determined in equilibrium by the temperature $T$
and the coupling strength $\lambda$.
Therefore we can conclude, that the generalized transport equation
(\ref{eq:general_transport}) is sufficient to describe the
correct relaxation of the chemical potential $\mu$.

\begin{figure}[t]
\centerline{\psfig{file=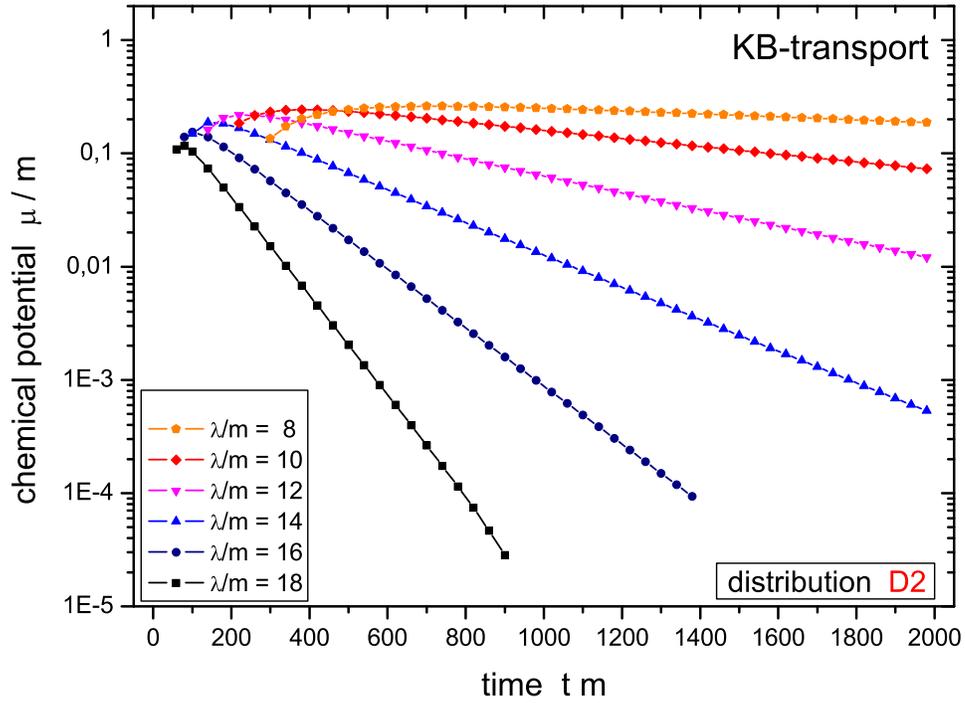,width=15cm}} \vspace{1.0cm}
\caption{\label{fig:4phd_gradient_muall} Relaxation of the
chemical potential $\mu / m$ for initial distribution D2 with
coupling constants $\lambda / m = $ 8, 10, 12, 14, 16, 18 using
the generalized transport equation (\ref{eq:general_transport}).}
\end{figure}

Finally, we study the relaxation of the chemical potential
as a function of the coupling strength $\lambda$.
To this aim we display in Fig. \ref{fig:4phd_gradient_muall}
the relaxation of the chemical potential $\mu$ for the initial
distribution D2 for coupling constants
$\lambda / m = $ 8, 10, 12, 14, 16, 18 as obtained form the
generalized transport equation (\ref{eq:general_transport}).
For all coupling constants $\lambda$ the chemical potential $\mu$ is
reduced exponentially in time $\propto \exp(- \Gamma_{\mu} \cdot t)$
and thus allows for the determination of
a proper relaxation rate $\Gamma_{\mu}$.

\begin{figure}[th]
\centerline{\psfig{file=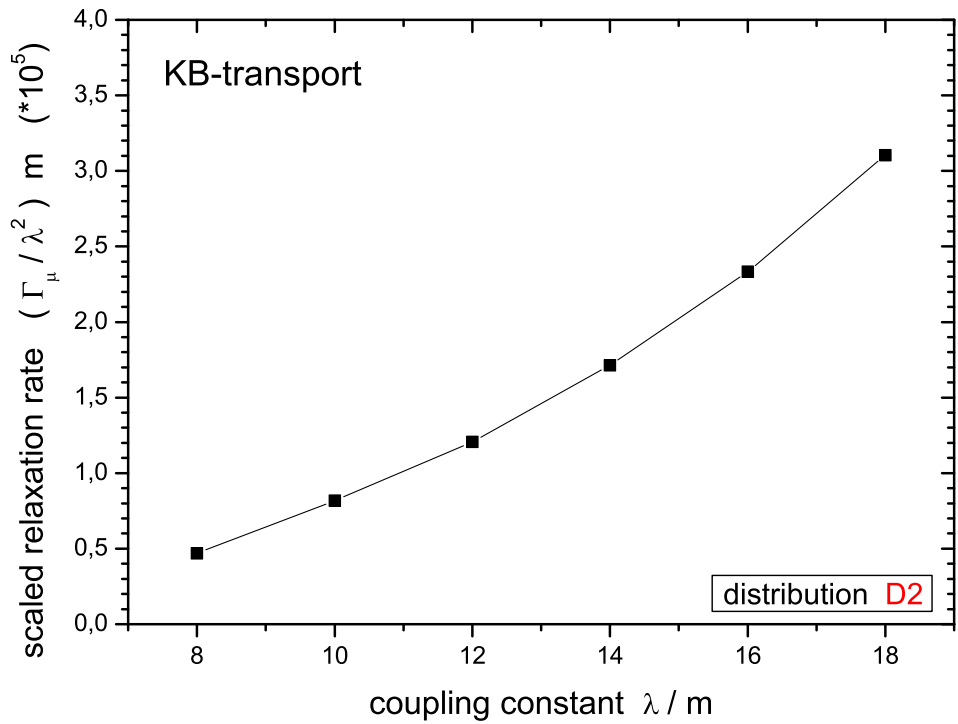,width=15cm}} \vspace{1.0cm}
\caption{\label{fig:4phd_gradient_murate} Scaled relaxation rate
$\Gamma_{\mu} / \lambda^2$ of the chemical potential as a function
of the coupling strength $\lambda / m$ for the initial
distribution D2.}
\end{figure}

The results for the relaxation rates $\Gamma_{\mu}$ of the
chemical potential as a function of the coupling strength
$\lambda$ are displayed in Fig. \ref{fig:4phd_gradient_murate}.
They are scaled by the coupling constant squared ($\lambda^2$) in
order to take into account the overall coupling dependence of the
collisional self-energy. We see that the relaxation rate
$\Gamma_{\mu}$ increases much stronger than quadratically with the
coupling strength. Whereas the relaxation is very weak for small
and medium couplings $\lambda / m \leq$ 8, it increases
considerably for larger interaction strength. By inspection of the
analytical estimate for the relaxation rate of the chemical
potential (Eq. (4.28) in Ref. \cite{Juchem03}) we find an
explanation for this strong dependence: At first here an overall
factor of the coupling constant squared ($\lambda^2$) enters the
expression for the relaxation rate, that stems from the
collisional integral in terms of the scattering (sunset)
self-energies. This factor, only, would yield a constant line in
Fig. \ref{fig:4phd_gradient_murate} and thus underestimate the
observed behaviour significantly. Therefore, one has to keep in
mind the additional $\lambda$ dependence of the (equilibrium)
spectral function that strongly influences the estimate for the
relaxation rate since it appears in the energy-momentum
integration weights several times. Thus the relaxation rate is
determined explicitly via the $K_2$-term (Eq. (4.26) in Ref.
\cite{Juchem03}) and implicitly through the spectral functions in
both contributions, $K_1$ and $K_2$ (Eqs. (4.26) in Ref.
\cite{Juchem03}), by the coupling strength $\lambda$ in a
nonlinear way.

We conclude that -- though there is a small relative shift of the
different time scales of kinetic and chemical equilibration as a
function of the coupling strength $\lambda$ with respect to the
full KB solutions -- the results of the generalized transport
equations are very similar. The differences we attribute to higher
order multi-particle effects in off-shell transitions. While the
kinetic equilibration proceeds approximately with the coupling
constant squared (as indicated by the calculations for
non-polar-symmetric systems), the chemical relaxation rate is a
higher order process in $\lambda$ as seen from Fig.
\ref{fig:4phd_gradient_murate}. Thus the chemical equilibration
moves with increasing coupling strength $\lambda$ to earlier times
relative to the kinetic relaxation, which is governed by terms
$\sim \lambda^2$ (see next Subsection).

\subsection{\label{sec:gradient_quadrupole} Quadrupole Relaxation}

\begin{figure}[th]
\centerline{\psfig{file=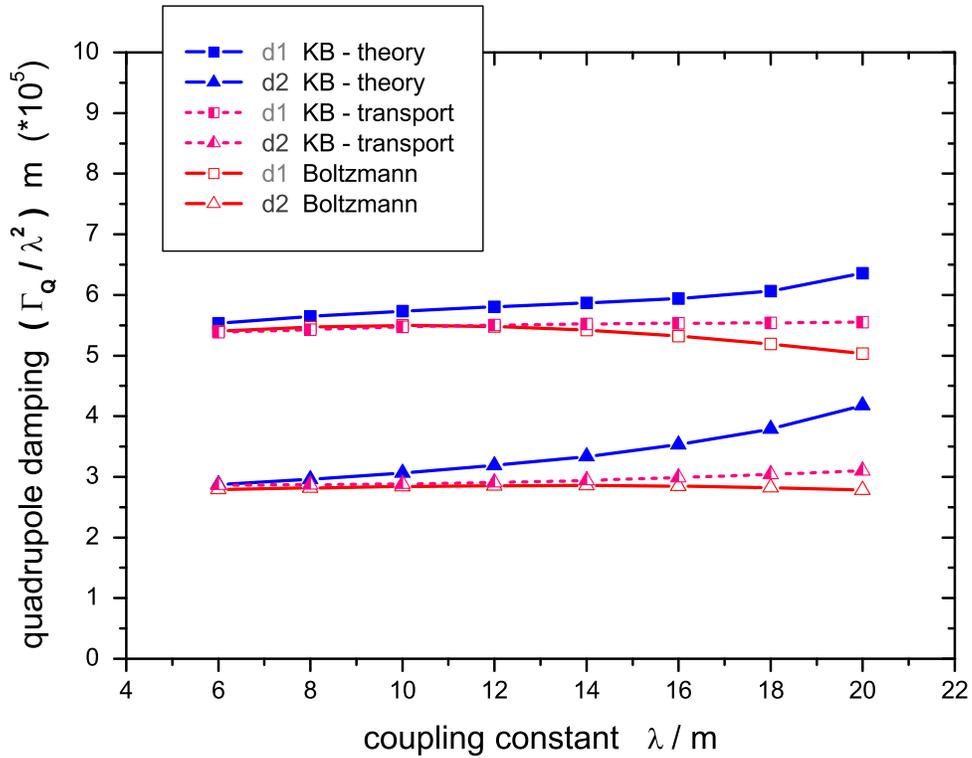,width=15cm}} \vspace{1.0cm}
\caption{\label{fig:4phd_gradient_quad} Scaled Relaxation rate for
the generalized transport equation as a function of the coupling
strength (half-filled symbols) for the initial distributions d1
(squares) and d2 (triangles). Additionally the results obtained
within the Kadanoff-Baym (full symbols) and the Boltzmann
calculation (open symbols) are shown for comparison. }
\end{figure}

In this Section we no longer restrict to polar symmetric systems
and discuss the time evolution of more general initial distributions
within the generalized transport approximation (\ref{eq:general_transport}).
We start with conditions similar to those employed in
Section V of Ref. \cite{Juchem03}, but combined with the initialization scheme
for the semiclassical limit.
Again -- as in the full Kadanoff-Baym and the Boltzmann case (Section V of Ref. \cite{Juchem03}) --
the decrease of the quadrupole moment of the distribution
\bea
\label{quadpole}
Q(t) \: = \:
\frac{\displaystyle{\int \!\! \frac{d^{2}p}{(2 \pi)^2} \;\;
[\, p_x^2 - p_y^2 \,] \;\; N(\vec{p},t)}}
     {\displaystyle{\int \!\! \frac{d^{2}p}{(2 \pi)^2} \;\;
N(\vec{p},t)}} \; ,
\eea\\
is approximately exponential in time ($\propto \exp(-\Gamma_Q \cdot t)$) and
thus allows for the extraction of a quadrupole damping rate $\Gamma_Q$.
The scaled quadrupole damping rates -- as obtained for the two
initial distributions d1 and d2 (cf. Section V of Ref. \cite{Juchem03})
-- are displayed in
Fig. \ref{fig:4phd_gradient_quad}
as a function of the coupling strength $\lambda / m$.
The calculations show that the quadrupole relaxation rates within the
semiclassical approximation (\ref{eq:general_transport})
for both initial distributions d1 and d2 is well within in the range
of the full Kadanoff-Baym and the on-shell Boltzmann case.
Additionally, the quadrupole relaxation rate is rather flat
in the coupling $\lambda$ when divided by the coupling constant
squared ($\lambda^2$) as already observed for the other two evolution
schemes in Section V of Ref. \cite{Juchem03}. Nevertheless, the
relaxation in the full KB-theory (\ref{kabaeqcs}) proceeds
slightly faster than in the transport limits for large couplings.
The latter effect is again attributed to higher order off-shell
transition effects which are no longer incorporated in the generalized
transport equation.

\subsection{\label{sec:gradient_validity}
Validity of the Gradient Approximation}

As we have seen in the previous Subsections
the generalized transport equation (\ref{eq:general_transport})
leads to a good agreement with the Kadanoff-Baym dynamics.
This indicates that the semiclassical limit can be applied
without loosing essential features of the full quantum dynamics for
homogeneous systems (in particular for the initializations D1, D2 and D3).
We recall, that the underlying assumption for the validity of the
first order gradient expansion scheme -- which has been used to derive
the generalized transport equation -- is that all functions are slowly
evolving in the mean space and time coordinates.
Thus, in comparison to the first order time derivatives, the second
order time derivatives should be small, such that they
can be neglected to a good approximation.
In this Subsection we will study now this criterion in a
more quantitative way.

\begin{figure}[htp]
\centerline{\psfig{file=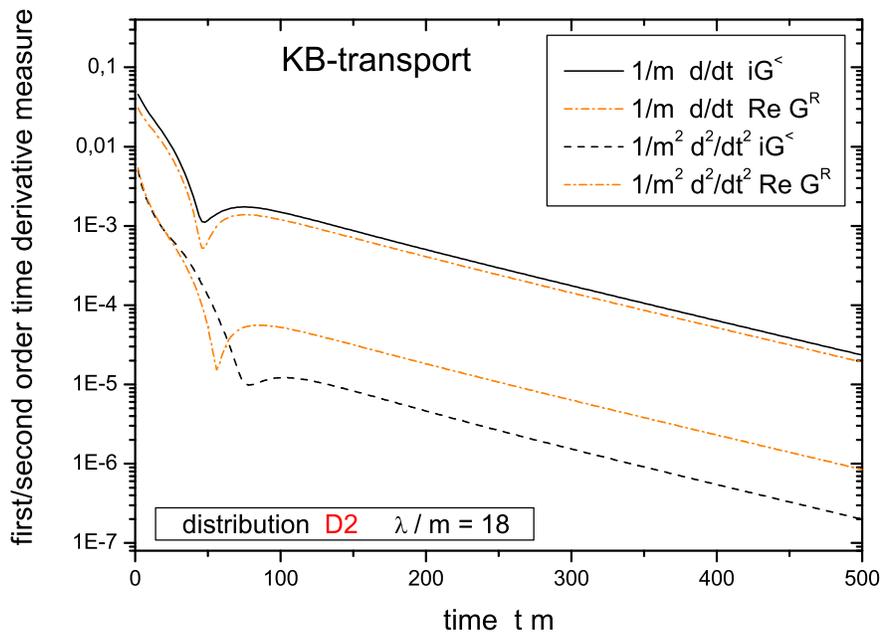,width=13.5cm}} \vspace{-1.0cm}
\caption{\label{fig:4phd_gradient_gradch_g} Time evolution of the
first/second order time derivative measure for the Green functions
$i\bar{G}^{<}$ and $Re\,\bar{G}^{R}$ for initial distribution D2
with coupling constant $\lambda / m =$ 18. During the whole
evolution the second order time derivatives are more than an order
of magnitude smaller than the first order derivatives supporting
the validity of the gradient expansion. Note that time derivatives
in our convention are derivatives with respect to the
dimensionless quantity $tm$. }
 \end{figure}

\begin{figure}[htp]
\vspace{-0.0cm} \centerline{\psfig{file=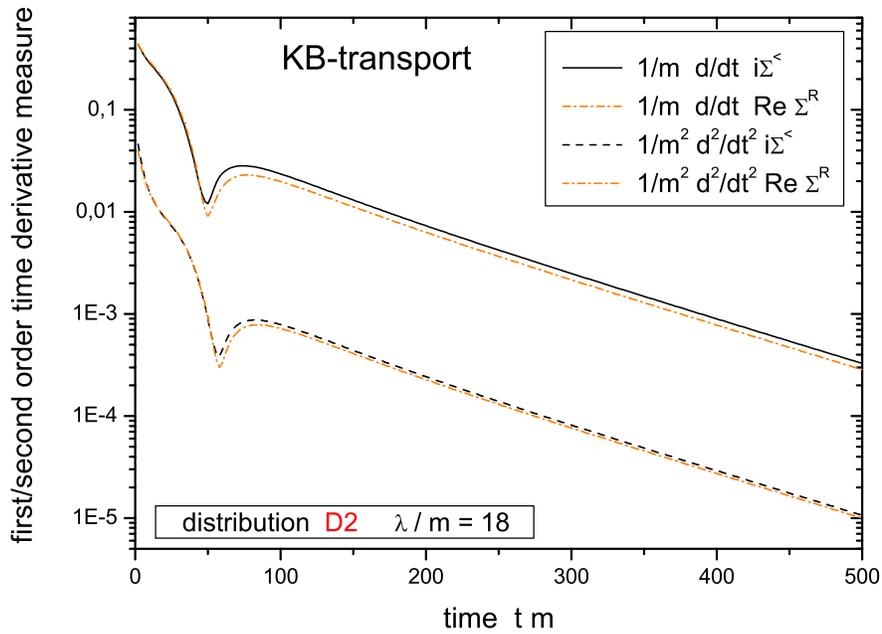,width=13.5cm}}
\vspace{-1.0cm} \caption{\label{fig:4phd_gradient_gradch_s} Time
evolution of the first/second order time derivative measure for
the functions $i\bar{\Sigma}^{<}$ and $Re\,\bar{\Sigma}^{R}$ for
initial distribution D2 with coupling constant $\lambda / m =$ 18.
During the whole evolution the second order time derivatives are
significantly smaller than the first order contributions. Note
that time derivatives in our convention are derivatives with
respect to the dimensionless quantity $tm$.}
\end{figure}

To this aim we consider as a relative measure the energy-momentum
integrals over the absolute value of first and second order time derivatives
of various functions entering the generalized transport equation.
Explicitly this measure is given at time $t$ by \\
\bea
M^{1/2}_{F}(t) \; =\;
\int \frac{d^{D}p}{(2\pi)^{D}} \;\;
\left| \; \partial_t^{(1/2)} \; \bar{F}(\vec{p},p_0,t) \; \right|
\eea\\
for an arbitrary function $\bar{F}$ in Wigner space.
In the following we take into account the time
derivatives of the Green functions and the self-energies,
i.e. \\
\bea
\bar{F} \: \in \: \{ \, i\bar{G}^{<},\,i\bar{\Sigma}^{<},\,
Re\,\bar{G}^{R},\,Re\,\bar{\Sigma}^{R} \,\} \: .
\eea\\
The time evolution of these measures is shown in
Fig. \ref{fig:4phd_gradient_gradch_g} for the first and
the second order time derivatives of the Green function
and in Fig. \ref{fig:4phd_gradient_gradch_s} for the collisional
and the retarded self-energies.
The calculation has been performed for the initial distribution D2
with a coupling strength of $\lambda / m =$ 18.
For the Green functions as well as the self-energies the
second order time derivatives are about one order of magnitude
smaller than the first order expressions.
Thus the underlying assumption of the first order gradient
expansion is fulfilled very well indicating that the results
obtained within the semiclassical scheme should match with those
for the full quantum evolution to a large extent.
This is exactly what we have found from the explicit comparison
of the evolution of the equal-time Green functions as well as the
spectral functions in Sections
\ref{sec:gradient_equilibration} and
\ref{sec:gradient_spectralevolution}.

\subsection{\label{sec:gradient_boma} Generalized Transport in
Botermans-Malfliet Form}

In this Subection we will perform a comparison of the generalized
transport equation in the original {Kadanoff-Baym} (KB) form
(\ref{eq:general_transport}) with the modified
{Botermans-Malfliet} (BM) form (\ref{eq:general_transport_bm}). As
discussed in detail in Section 2.2 the latter form results from
the replacement of the collisional self-energy by
\begin{equation} \label{replace}
i\bar{\Sigma}^{<} \rightarrow i\bar{G}^{<} \cdot \bar{\Gamma} / \bar{A}
\end{equation}
 in the second Poisson bracket on the l.h.s. of the original
kinetic equation (\ref{eq:general_transport}). This replacement
leads to a consistent first order equation in the gradients and
achieves consistency of the resulting transport equation with the
corresponding generalized mass-shell relation
(\ref{eq:general_mass}).

\begin{figure}[th]
\centerline{\psfig{file=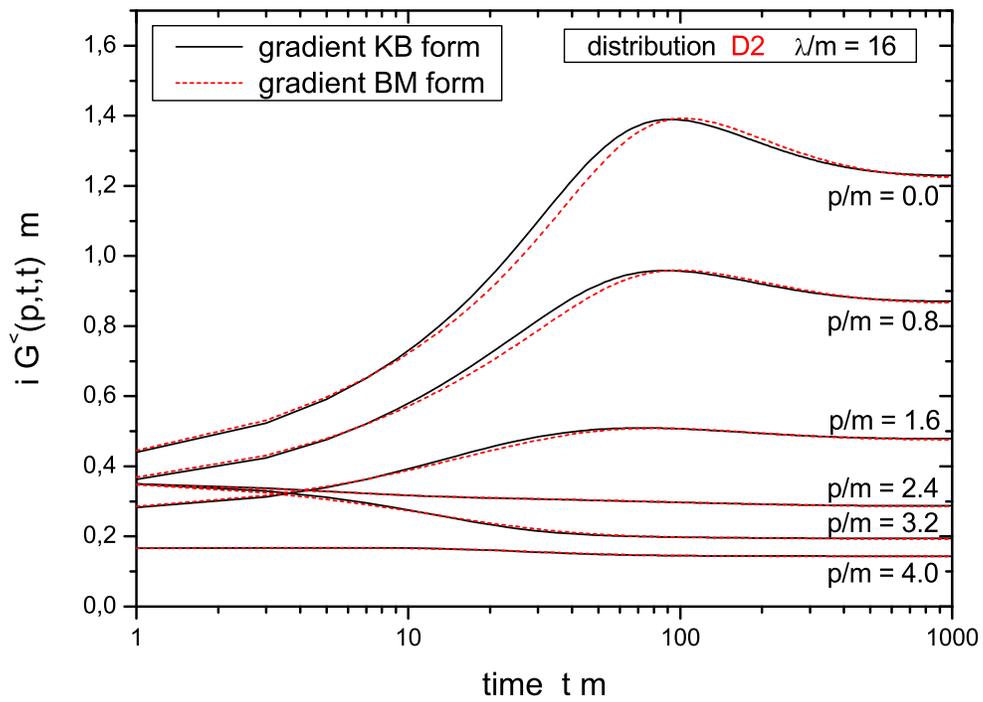,width=15cm}} \vspace{-1.0cm}
\caption{\label{fig:4phd_gradient_bm} Time evolution of various
equal-time Green function momentum modes within the generalized
transport equation (original KB form, solid lines) and within the
Botermans-Malfiet (BM) form (dashed lines) for initial
distribution D2 with coupling constant $\lambda/m = 16$.}
\end{figure}

In Fig. \ref{fig:4phd_gradient_bm} we compare the time evolution
within the generalized transport equation in the KB form to the
consistent equation in BM form. In this respect several momentum
modes of the equal-time Green function are displayed evolving in
time from an initial distribution D2 for a coupling constant
$\lambda / m =$ 16. We find that the deviations between both
approximations (KB and BM) are rather moderate. Only for very
small momentum modes $|\,\vec{p}\,| / m \leq$ 1.6 deviations
between both calculational modes are visible. For the very low
momentum modes the range of difference starts at $t \cdot m
\approx$ 10 and extends to $t \cdot m \approx$ 100 for the
non-zero modes. For the zero momentum mode the deviation lasts
even longer. In this region the semiclassical transport in the BM
form is slightly 'slower' than in the original KB choice.
Nevertheless, also the BM form exhibits the typical overshooting
behaviour of the low momentum modes beyond the stationary limit as
observed for the KB form. However, the maxima are shifted slightly
to later times. Finally, both gradient approximations converge in
the long-time limit to very similar configurations.

\begin{figure}[th]
\centerline{\psfig{file=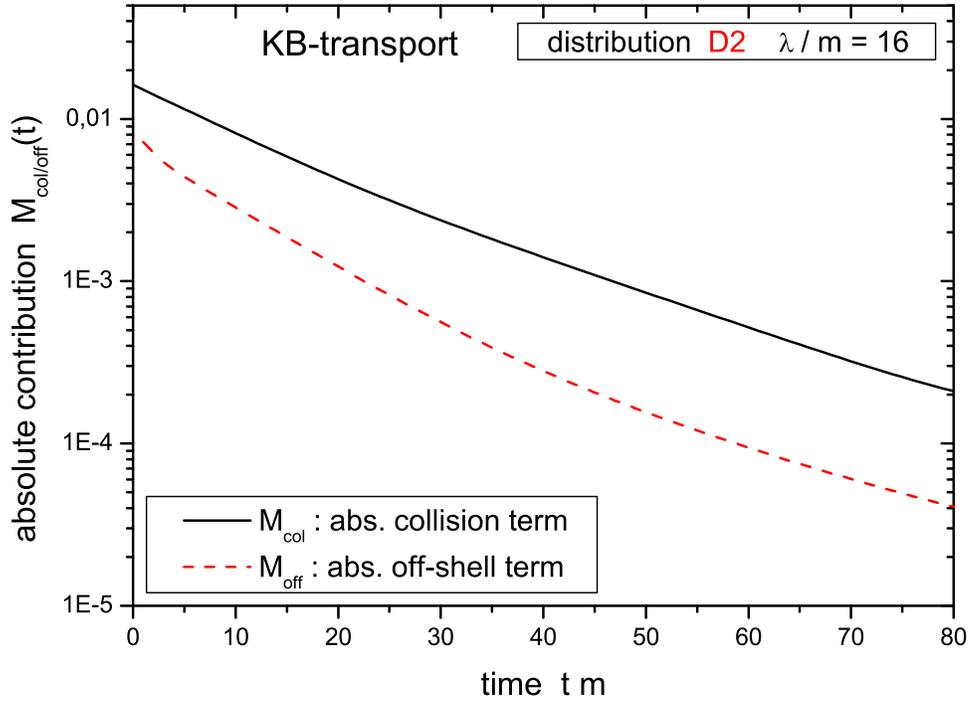,width=15cm}} \vspace{1.0cm}
\caption{\label{fig:4phd_gradient_abs} Time evolution of the
measure functions for the collision term (\ref{eq:gradient_mcol})
and the difference of the second Poisson bracket in standard (KB)
and consistent (BM) form (\ref{eq:gradient_moff}) for the initial
distribution D2 with coupling constant $\lambda / m =$ 16. Both
contributions decrease in time in accordance with the
equilibration of the system. The Poisson contribution (dashed
line) is always significantly smaller than that of the collision
term (solid line).}
\end{figure}

As the final part of the comparison we investigate the
approximation (\ref{replace})
quantitatively and
 introduce a measure for the absolute strength of the
collision term by integrating the collision rates over energy and momentum.
Since the collision term vanishes, if the substitution (\ref{replace}) holds
exactly, i.e. in equilibrium, the absolute size gives an
idea about the validity of this replacement at the zero
order level.
Explicitly the measure reads: \\
\bea
\label{eq:gradient_mcol}
M_{col}(t) & = & \int \!\! \frac{d^{D}p}{(2\pi)^{D}} \;\;
\frac{1}{|\,2 p_0\,|} \:
\left| \;
\phantom{\frac{1}{2}} \!\!
i\bar{\Sigma}^{<} \; i\bar{G}^{>}
\; - \;
i\bar{\Sigma}^{>} \; i\bar{G}^{<}
\; \right| \; .
\eea\\
Furthermore, we define a measure for the deviation of the
second Poisson bracket in the original and in the
consistent Botermans-Malfliet formulation.
It is given by the integration over the absolute differences
of the Poisson terms as \\
\bea
\label{eq:gradient_moff}
M_{off}(t) & = & \int \!\! \frac{d^{D}p}{(2\pi)^{D}} \;\;
\frac{1}{|\, 2 p_0 \,|} \:
\left| \;
\{ \, i\bar{\Sigma}^{<} \, , \,  Re\,\bar{G}^{R} \, \}
\; - \;
\{ \, i\bar{G}^{<} \, \frac{\bar{\Gamma}}{\bar{A}} \, , \,
Re\,\bar{G}^{R} \, \}
\; \right| \; .
\eea\\
In Fig. \ref{fig:4phd_gradient_abs} we display the
time evolution of the measure (\ref{eq:gradient_mcol})
(solid line) and the measure (\ref{eq:gradient_moff})
(dashed line).
The calculation has been performed for the initial state D2
with a coupling constant $\lambda / m =$ 16.
We find that both measures decrease as a function of time
in accordance with the equilibration of the system.
Additionally, the contribution from the difference of the
Poisson brackets (\ref{eq:gradient_moff}) is always smaller
than the one stemming from the collision term
(\ref{eq:gradient_mcol}).
This indicates that the replacement of the collisional self-energy
$i\bar{\Sigma}^{<} \rightarrow i\bar{G}^{<} \cdot \bar{\Gamma} / \bar{A}$
is more reliable when it takes place at the gradient level
in accordance with the assumption of Botermans and Malfliet.
However, the relative suppression is not very large.

In summary we point out that the approximation of
the full Kadanoff-Baym dynamics by the generalized transport
equations in Kadanoff-Baym (\ref{eq:general_transport}) or
Botermans-Malfliet form (\ref{eq:general_transport_bm}) holds very
well for the different momentum modes of the Green function
$i\bar{G}^{<}$ itself. Slight deviations are only visible for the
zero momentum mode at early to intermediate times (Figs.
 \ref{fig:4phd_gradient_modes} and
\ref{fig:4phd_gradient_bm}) for a logarithmic representation of
the time axis. Consequently, the characteristic features of
quantum equilibration obtained for the full Kadanoff-Baym theory
are retained in the generalized transport limits. The validity of
these transport equations -- based on a first order gradient
expansion -- could be shown explicitly, since second order
gradient terms turned out to be smaller by more than an order of
magnitude (cf. Figs. \ref{fig:4phd_gradient_gradch_g} and
\ref{fig:4phd_gradient_gradch_s}).

\section{Summary and Outlook}
In this work we have studied the quantum time evolution of $\phi^4$-field
theory for homogeneous systems in 2+1 space-time dimensions for
far-from-equilibrium initial conditions in extension of our studies in Ref.
\cite{Juchem03}. The
three-loop approximation for the CTP 2PI effective action has been
employed, i.e. the tadpole and sunset self-energies.
The tadpole contribution corresponds to a
dynamical mass term whereas the sunset self-energy is responsible
for dissipation and an equilibration of the system. Since both
self-energies are ultraviolet divergent, they had to be
renormalized by including proper counter terms. The numerical solutions for different initial
configurations out of equilibrium (with the same energy density)
show, that the asymptotic state achieved for $t \rightarrow
\infty$ is the same for all initial conditions in the full
Kadanoff-Baym theory.

In order to improve the standard on-shell (or quasiparticle)
Boltzmann transport theory we have derived generalized off-shell transport
equations from the Kadanoff-Baym equations in phase-space
representation by restricting to first order derivatives in $x$
and $p$. In fact, it could been shown explicitly, that second
order derivatives (of self-energies and Green functions) are
smaller than the first order derivatives by at least an order of
magnitude. As a consequence the dynamics within the generalized
transport formulation shows a very similar structure as the full
quantum solution. This is clearly seen for the propagation of the
equal-time momentum modes of the Green functions and the evolution
of the spectral function for all configurations considered. It
includes, in particular, the overshooting behaviour of the low
momentum equal-time modes, which occurs at intermediate times
depending on the initial distribution. Furthermore, the evolution
within the generalized transport equation leads to a stationary
state in the long time limit, which exhibits a full off-shell
equilibration with vanishing chemical potential. Even the
relaxation rates of the chemical potential obtained from the
semiclassical evolution agree very well with those of the full
Kadanoff-Baym theory. The dependence of the chemical relaxation rate on the
coupling constant is rather non-trivial, since it is strongly
affected by the equilibrium spectral function.  Thus we conclude
that the inclusion of the
dynamical spectral function -- as inherent in the semiclassical
approximation of the KB equations -- surpasses the shortcomings of
the on-shell Boltzmann limit.

Moreover, we have shown that the generalized transport equation in
Kadanoff-Baym (KB) form (\ref{eq:general_transport}) and in
Botermans-Malfliet (BM) form (\ref{eq:general_transport_bm}) lead
to comparable results for the time evolution of the initial
configurations considered. Only for the time-dependent occupation
of low momentum modes slight differences have been observed. This
is a typical quantum phenomenon related to the large de Broglie
wavelength of the low momentum modes. We stress that the
generalized transport equation in BM form can be used for the
off-shell description of realistic heavy-ion collisions
\cite{caju1,caju2,caju3,cas03} -- as well as many other
interacting systems in various fields of physics -- since it
allows for an economic solution within an extended testparticle
ansatz.

Coming back to the questions raised in the Introduction, we have
to answer the first one by 'probably no' since off-shell quantum
transitions in comparison to on-shell two-body scattering only
lead to slightly smaller relaxation times e.g. for the quadrupole
moment in momentum space. On the other hand, the off-shell
transitions are important (in case of the $\phi^4$-theory) for
chemical equilibration, which essentially proceeds via $1 \leftrightarrow 3$
transitions \cite{Juchem03}. The latter are forbidden in the on-shell
quasiparticle limit due to energy and momentum conservation. We recall that also
$0 \leftrightarrow 4$ processes are incorporated in the full
Kadanoff-Baym theory though their effect was found to be very limited
for the dynamical configurations investigated here.

Nevertheless, this still leaves us with the open question on the
relative importance of many-body scattering processes  $n
\leftrightarrow m$ for $n \geq 2, m > 2$. Here, an appropriate
formulation of $n \leftrightarrow m$ off-shell and on-shell
scattering processes has already been given in Ref.
\cite{Casspbar} as well as an application to the problem of
antibaryon annihilation and recreation by several mesons in
relativistic nucleus-nucleus collisions. A related study on the
problem of $2 \leftrightarrow 3$ transitions (e.g. bremsstrahlung
for massless particles) on the parton level will be presented in
the near future \cite{Xu}.

\section*{Acknowledgements}
The authors acknowledge inspiring discussions with J. Knoll, S.
Leupold and K. Morawetz.

%--------------------------------------------------------------------

%
%
%

\newpage


\newpage

\end{document}